\documentclass[twocolumn,floatfix, superscriptaddress]{revtex4-2}
\usepackage[utf8]{inputenc}

\usepackage{comment}
\usepackage{xcolor}
\usepackage{amsmath}
\usepackage{amsfonts}
\usepackage{float}
\usepackage{natbib}
\usepackage{graphicx}
\usepackage[caption=false]{subfig}
\usepackage{lipsum}
\usepackage{placeins}
\usepackage{booktabs}
\usepackage{hyperref}
\hypersetup{
    colorlinks,
    linkcolor={blue!50!black},
    citecolor={blue!50!black},
    urlcolor={blue!80!black}
}

\begin{document}

\title{Investigation of the effects of non-Gaussian noise transients and their mitigation in parameterized gravitational-wave tests of general relativity}

\author{Jack Y.\ L.\ Kwok} 
\email[Email:\ ]{jackkwok@link.cuhk.edu.hk}
\affiliation{Department of Physics, The Chinese University of Hong Kong, Shatin, N.T., Hong Kong}
\author{Rico K.\ L.\ Lo}
\author{Alan J.\ Weinstein}
\affiliation{ LIGO, California Institute of Technology, Pasadena, California 91125, USA}
\author{Tjonnie G.\ F.\ Li} 
\affiliation{Department of Physics, The Chinese University of Hong Kong, Shatin, N.T., Hong Kong}
\affiliation{Institute for Theoretical Physics, KU Leuven, Celestijnenlaan 200D, B-3001 Leuven, Belgium}
\affiliation{Department of Electrical Engineering (ESAT), KU Leuven, Kasteelpark Arenberg 10, B-3001 Leuven, Belgium}
\date{\today}

\begin{abstract}
	The detection of gravitational waves from compact binary coalescence by Advanced LIGO and Advanced Virgo provides an opportunity to study the strong-field, highly relativistic regime of gravity. Gravitational-wave tests of general relativity (GR) typically assume Gaussian and stationary detector noise and, thus, do not account for non-Gaussian, transient noise features (glitches).
	We present the results obtained by performing parametrized gravitational-wave tests on simulated signals from binary-black-hole coalescence overlapped with three classes of frequently occurring instrumental glitches with distinctly different morphologies. 
We then review and apply three glitch mitigation methods and evaluate their effects on reducing false deviations from GR.
	By considering nine cases of glitches overlapping with simulated signals, we show that the short-duration, broadband blip and tomte glitches under consideration introduce false violations of GR, and using an inpainting filter and glitch model subtraction can consistently eliminate such false violations without introducing additional effects.
\end{abstract}

\maketitle

\section{Introduction}
Over a century after its formulation in 1915, Einstein's general relativity (GR) remains as the accepted theory of gravity, passing all precision tests to date \cite{tgr_review}.
In the weak-field, slow-motion regime, where the effects of metric theories of gravity can be approximated as higher-order \emph{post-Newtonian} (PN) corrections to the Newtonian theory \cite{gravitation}, GR lies within the stringent bounds set by solar-system tests and pulsar tests \cite{will2006,will2014}.
Recent attention has turned to testing GR in the strong-field, highly relativistic regime \cite{will2006},
which potentially suggests high-energy corrections to the Einstein-Hilbert action \cite{stelle}, making GR compatible with standard quantum field theory \cite{tgr_review}.
One approach of probing the strong-field regime is through the detection of gravitational waves (GWs),
which carry information about its astrophysical origin \cite{schutz_review}.

Of all strong-field astrophysical events that could be probed using GWs, the \emph{coalescence} of stellar-mass binary black holes (BBHs), which can be schematically divided into \emph{inspiral}, \emph{merger} and \emph{ringdown} (IMR) stages, plays a crucial role in testing GR \cite{tgr_review}. Since the orbital separation of BBHs can reach far below the last stable orbit before merging, the generated gravitational field can be many orders of magnitudes stronger than other astrophysical events observed so far \cite{yunes_review, yunes, tigerimr, pnp1,pnp2,pnp3,pnp4}.
Moreover, GWs emitted by coalescing BBHs offer one of the cleanest tests of GR, as matter and electromagnetic fields are negligible for most sources \cite{enveffect, yunes}, and the emitted GWs essentially propagate through matter unimpeded \cite{yunes}, enabling precision tests of the strong-field dynamics of GR.
Since 2015, Advanced LIGO \cite{LIGO} and Advanced Virgo \cite{VIRGO} have jointly announced over 40 confident detections of GWs from coalescing BBHs \cite{gwtc, gwtc2, gwtc2_1}. 

Several GW tests of GR using coalescing BBHs are developed to test for \emph{generic} deviations from GR without the need for signal models from competing theories of gravity \cite{yunes}. For example, consistency tests search for excess power in the residual noise after subtracting a best-fit GR waveform \cite{tgr150914} or compare the source parameters inferred using only high-frequency data to that inferred using only low-frequency data \cite{tgr150914};
parametrized tests introduce parametrized deformations to waveform approximations to GR and infer the extent of deviation using Bayesian parameter estimation \cite{tigerimr}.
To date, no evidence for violations of GR has been identified using GWs emitted by coalescing BBHs \cite{tgrgwtc,o3a_tgr}.

\begin{figure}[t]
	\includegraphics[width=\linewidth]{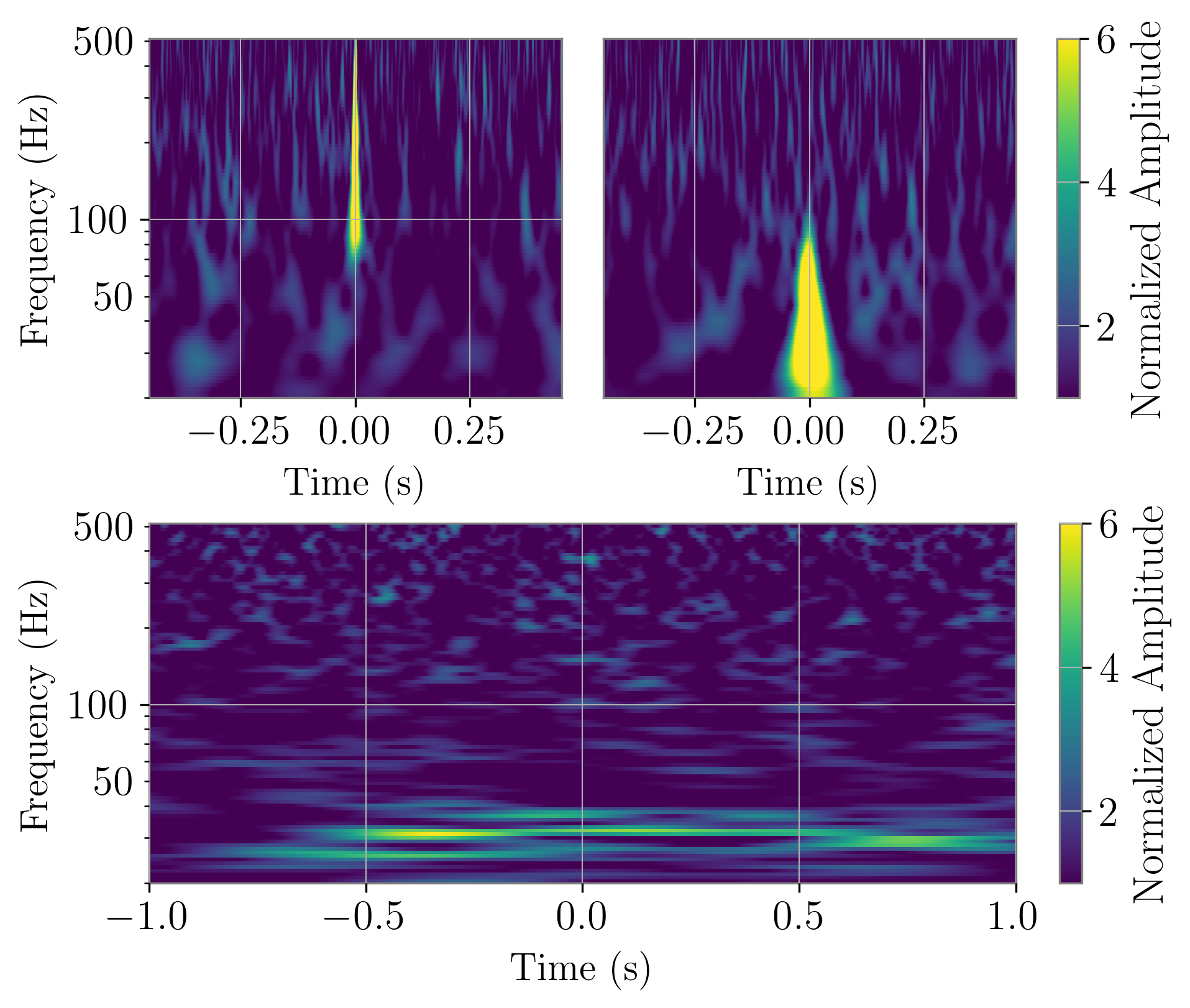}
	\caption{Glitches with similar morphology can be algorithmically categorized into different classes \cite{glitch2}.
		A time-frequency representation, called a \emph{Q scan} (or \emph{Omega scan}) \cite{chatterji}, where the duration of each time-frequency bins varies inversely with frequency and linearly with a parameter $Q$, is commonly used to visualize glitches \cite{glitch2, omicron}.
		$Q$ scans of three frequently occurring glitches (top left, blip; top-right, tomte; bottom, scattered light) during the O3 observing run are shown. The value of $Q$ used is 8, 8, and 40, respectively.
		The color represents the normalized amplitude (square root of the normalized power) in each time-frequency bin.
}
\label{fig:glitch}
\end{figure}

Aside from GWs, output from GW detectors is attributed to many independent sources of random noise \cite{saulson}. Assuming that noise characteristics remain stationary over observation timescales, detector noise is typically modeled as stationary and Gaussian in GW data analysis in light of the central limit theorem \cite{lalsuite, veitch2015parameter}.
However, these assumptions cannot account for transient, non-Gaussian noise features,
commonly referred to as \emph{glitches}  \cite{glitch1, glitch2,glitch3}.
Glitches pose significant problems to GW searches \cite{glitch2} and may bias GW data analysis by violating the noise model.
Three glitches from commonly seen glitch classes during the O3 observing run are shown in Fig.\ \ref{fig:glitch}.
 
Many efforts are made to identify and classify glitches \cite{veto1, veto2, veto3, veto4, omicron, glitch2}.
Once a glitch is identified, the data containing the glitch can be removed using various mitigation methods  \cite{gstlalgating, pycbcgating, bayeswave, bayeswave2, inpaint}.
The effects of glitches and their mitigation on the inference of source parameters have been studied in the context of glitches similar to that affecting GW170817 \cite{gw170817_mitigation}.
It is of interest to extend the study to parametrized tests of GR, as the additional degree(s) of freedom introduced by parametrized deformations of the signal model may enhance such effects.

This article is structured as follows: Section \ref{sec2} describes the typical data model used in GW data analyses \cite{lalsuite, veitch2015parameter}, which comprises of a GW signal in additive stationary and Gaussian noise.
Section \ref{parameterized_tests} introduces a parametrized test of GR involving the phase parametrization of an IMR waveform model \cite{imrphenomp1}.
Section \ref{glitch_mitigation} introduces three glitch mitigation methods to be applied in our investigation, namely bandpass filtering, inpainting, and glitch model subtraction.
Section \ref{sec:results} presents the results obtained by performing the parametrized test of GR to glitch-overlapped BBH-coalescence GW signals before and after glitch mitigation.

\section{Data Model} \label{sec2}
A GW detector is designed to respond linearly to the fractional change in arm length, or \emph{strain} \cite{saulson}. The time series of detector output data $\boldsymbol{d}$, sampled at time $t_k$ at constant sampling interval $\Delta t$, can thus be expressed as a linear superposition of a time series of the GW strain signal $\boldsymbol{h}$ and a time series of detector noise $\boldsymbol{n}$:
\begin{equation}
	\boldsymbol{d}(t_k) = \boldsymbol{h}(t_k) + \boldsymbol{n}(t_k) \; . \label{data_model}
\end{equation}
In Eq.\ (\ref{data_model}) and in subsequent discussion, boldface denotes the matrix representation of specified quantities.

\subsection{Stationary Gaussian noise model}
Assuming that a \emph{large} number of independent noise sources contribute linearly to the detector noise $\boldsymbol{n}$,
the central limit theorem states that the probability density distribution of the noise $\boldsymbol{n}$ tends to follow a multivariate \emph{Gaussian} distribution \cite{davenport}:
\begin{equation}
	P(\boldsymbol{n}) = \frac{1}{\sqrt{(2\pi)^N |\mathbf{\Sigma}|}} e^{-\frac{1}{2}(\boldsymbol{n}-\boldsymbol{\mu})^T \boldsymbol{\Sigma}^{-1} (\boldsymbol{n}-\boldsymbol{\mu})} \; , \label{multivariate}
\end{equation}
which is uniquely defined by the \emph{covariance matrix} $\Sigma_{ij} = E[(n(t_i)-\mu(t_i))(n(t_j)-\mu(t_j))]$ and the mean vector $\mu_i = E[n(t_i)]$, where $E[\cdot]$ and $\mid\cdot\mid$ denote the expectation and determinant operation, respectively.
The diagonal (off-diagonal) terms of the covariance matrix are the variances at each instance of time (correlations between data from different instances of time).

If the number of samples $N$ is large, it is undesirable to invert the $N\times N$ covariance matrix in Eq.\ (\ref{multivariate}). Instead, we consider the joint probability density in the frequency domain, which is also a multivariate Gaussian distribution \cite{davenport}.
With the assumption of stationarity---i.e.\ the joint probability density distribution is time invariant---the covariance matrix in the frequency domain is diagonalized in the infinite-duration limit \cite{romano}. This relation can be approximated for the finite-duration discretely sampled time series, giving the following approximation to the joint probability density in the frequency domain \cite{romano} (for even $N$), also known as the \emph{Whittle likelihood} \cite{whittle} in the context of statistical inference:
\begin{equation}
	P(\boldsymbol{n}) \simeq \prod^{N/2-1}_{j=0} \frac{2\Delta f}{\pi S_n(f_j)} \exp\left(-\Delta f \frac{2|\tilde{n}_j|^2}{S_n(f_j)} \right) \; \label{whittle} ,
\end{equation}
where $f_j \equiv j/N\Delta t$. The quantity $S_n(f_j)\equiv 2|\tilde{n}(f_j)|^2 /T$ is scaled from the diagonal terms of the covariance matrix in the frequency domain, $\Delta f \equiv 1 /T$ is the \emph{frequency resolution} and the tilde denotes a discrete Fourier transformed (DFT) quantity:
\begin{equation}
	\tilde{n}_j \equiv \Delta t\ \mathrm{DFT}[n(t_k)] = \Delta t \sum^{N-1}_{k=0} n(t_k) e^{-2\pi ijk/N} \; .
\end{equation}

To motivate the quantity $S_n(f_j)$, called the \emph{one-sided power spectral density} (PSD), we invoke Parseval's theorem \cite{romano}:
\begin{equation}
	\sum^{N/2-1}_{j=0} S_n(f_j) \Delta f \equiv \frac{2}{T} \sum^{N/2-1}_{j=0} |\tilde{n}(f_j)|^2 \Delta f = \frac{1}{T} \sum^{N-1}_{k=0} |n(t_k)|^2\Delta t \; \label{parseval} ,
\end{equation}
and note that the rightmost side of Eq.\ (\ref{parseval}) returns the \emph{power} of the time series. Since the time series is real, we have $\tilde{n}(f_j)=\tilde{n}^*(-f_j)$. Consequently, we can sample only the frequency bins from 0 Hz to up to the \emph{Nyquist frequency} $1/2\Delta t$ and introduce the factor of 2 in Eqs.\ (\ref{whittle}) and (\ref{parseval}).

\subsection{Signal model} \label{waveform_model}
Since the two-body self-gravitating problem cannot be solved analytically in GR,
we generate simulated GW strain signals from coalescing BBHs using the frequency-domain precessing IMR waveform model \texttt{IMRPhenomPv2} \cite{imrphenomp1} in virtue of its good match with numerical relativity (NR) waveforms \cite{imrphenompv3} and low computational costs.

\begin{figure}
	\centering
	\includegraphics[width=\linewidth]{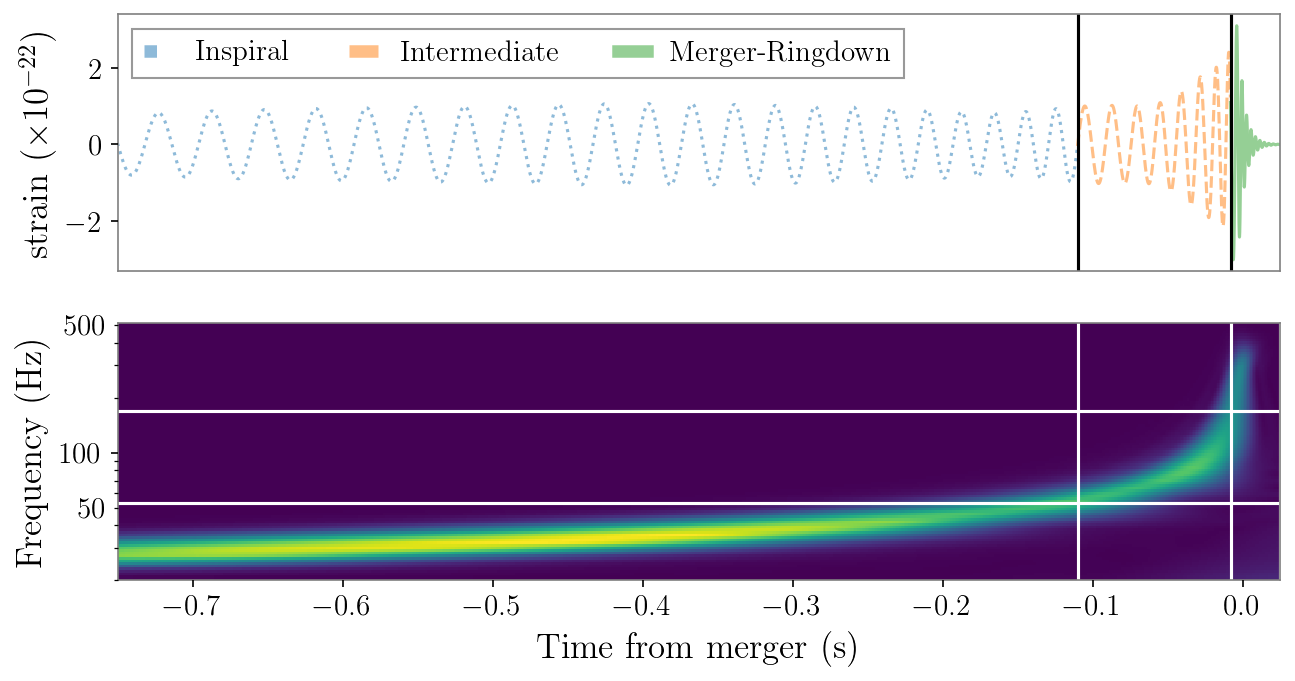}
	\caption{An example \texttt{IMRPhenomPv2} time-domain GW waveform (upper figure) and the corresponding instantaneous frequency (lower figure) plotted against time. In natural units, the two horizontal lines in the lower figure correspond to the frequencies $0.018/M$ (lower line) and  $f_{\mathrm{RD}} /2$ (upper line), which defines the boundaries of the inspiral, intermediate, and merger-ringdown stages in the frequency domain. The corresponding time-domain boundaries (vertical lines) are determined as the times when the instantaneous frequency of the signal intersects with the frequency-domain boundaries.}
	\label{grid}
\end{figure}

\texttt{IMRPhenomPv2} is a phenomenological waveform model constructed by combining PN-like inspiral waveforms with NR-calibrated merger-ringdown ansatz \cite{imrphenomp3}.
In natural units, the \emph{inspiral} stage of \texttt{IMRPhenomPv2} is modeled up to $f \sim 0.018/M$, where $M$ is the total mass of the system. The region with $Mf \geq 0.018$ is subdivided into an \emph{intermediate} stage with $0.018\geq Mf\geq 0.5f_{\mathrm{RD}}$, which bridges the inspiral stage to the \emph{merger-ringdown} stage modeled above half the ringdown frequency $f_{\mathrm{RD}}$ \cite{imrphenomp3}. Figure \ref{grid} illustrates the stages of coalescence of an example \texttt{IMRPhenomPv2} GW strain and its frequency evolution over time.

The phase of \texttt{IMRPhenomPv2} consists of terms with known frequency dependence. The coefficients of these terms, denoted as the \emph{phase coefficients} $p_i$, are the subjects of parametrized tests of GR to be discussed in \mbox{Sec.\ \ref{parameterized_tests}}.
The phase coefficients $p_i$ and the orbital evolution of the BBH depend only on the masses and spin angular momentum vectors of the component black holes \cite{imrphenompv3}, denoted as the \emph{intrinsic} parameters.
The phase coefficients $p_i$ can be categorized into three groups, depending on the stages of coalescence in which they predominantly assert their effect on \cite{imrphenomp3, tigerimr}
(i) the \emph{inspiral} PN coefficients $\{\varphi_0,\dots,\varphi_5,\varphi_{5l},\varphi_6,\varphi_{6l},\varphi_7\}$ and phenomenological coefficients $\{\sigma_0,\dots,\sigma_4\}$;
(ii) the \emph{intermediate} phenomenological coefficients $\{\beta_0,...,\beta_3\}$;
(iii) the \emph{merger-ringdown} phenomenological and black hole perturbation theory coefficients $\{\alpha_0,\dots,\alpha_5\}$.

Seven additional \emph{extrinsic} parameters, including the sky location, luminosity distance, polarization angle of the source, and the spatial orientation and orbital phase of the system at a reference frequency and time, respectively, are needed to determine the response of the GW detectors.

\section{Parametrized Tests of GR} \label{parameterized_tests}
We will focus on a test of GR which tests for parametrized deviations from GR. It assumes the stationary Gaussian noise model. As such this test provides quantitative indicators of whether glitches can result in false deviations of GR and whether glitch mitigation will reduce or amplify them.

In the test, \emph{fractional} deviations $\delta p_i$, also known as \emph{dephasing coefficients}, are introduced to the \texttt{IMRPhenomPv2} phase coefficients $p_i$ \cite{tigerimr}:
\begin{equation}
	p_i \mapsto p_i [1+\delta p_i] \; .
\end{equation}
For the exceptional case where $p_i=0$, such as $\varphi_1$, an \emph{absolute} deviation is instead introduced \cite{tigerimr}.
In practice, we do not allow some of the \texttt{IMRPhenomPv2} phase coefficients to deviate from their prescribed values, as they have large uncertainties or are degenerate with other coefficients or physical parameters \cite{tigerimr}.
We therefore perform tests with the remaining 14 dephasing coefficients, henceforth denoted as the \emph{testing} dephasing coefficients \cite{tigerimr}:
\begin{align}
	\{\delta p_i\} = \{&\delta \varphi_0,\dots,\delta \varphi_4,\delta\varphi_{5l},\delta\varphi_6,\delta\varphi_{6l},\delta\varphi_7,\nonumber \\
	&\delta \beta_2, \delta\beta_3,\delta\alpha_2,\delta\alpha_3,\delta\alpha_4\} \; .
\end{align}
The frequency dependence of the testing parameters $\delta p_i$ is shown in Table \ref{fdependence} \cite{tgr150914, imrphenomp2}.

\begin{table}[]
	\caption{The frequency dependence of the \texttt{IMRPhenomPv2} testing parameters used in parametrized tests of GR. The table is reproduced from Table 1 of Ref.\ \cite{tgr150914}. The coefficients $a$ and $b$ in the $f$ dependence of  $\delta\alpha_4$ are functions of the component masses and spins \cite{imrphenomp3}.} \label{fdependence}
\begin{tabular}{lcl}\toprule\toprule
\begin{tabular}[c]{@{}l@{}}Stage of coalescence\end{tabular} & \begin{tabular}[c]{@{}c@{}}\phantom{pp} $\delta p_i$\phantom{ppp}\end{tabular} & $f$ dependence \\ \toprule
Inspiral & $\delta\varphi_0$ & $f^{-5 /3}$ \\
	 & $\delta\varphi_1$ & $f^{-4 /3}$ \\ & $\delta\varphi_2$ & $f^{-1}$ \\
	 & $\delta\varphi_3$ & $f^{-2 /3}$ \\ & $\delta\varphi_4$ & $f^{-1 /3}$ \\
	 & $\delta\varphi_{5l}$ & $\log f$ \\ & $\delta\varphi_6$ & $f^{1 / 3}$ \\
	 & $\delta\varphi_{6l}$ & $f^{1 / 3} \log f$ \\
	 & $\delta\varphi_7$ &  $f^{2 / 3}$\\ \midrule
Intermediate & $\delta\beta_2$ & $\log f$ \\
	     & $\delta\beta_3$  & $f^{-3}$ \\ \midrule
Merger-ringdown & $\delta\alpha_2$ &  $f^{-1}$ \\
& $\delta\alpha_3$ & $f^{3 / 4}$  \\
	 & $\delta\alpha_4$ & $\tan^{-1}(af+b)$ \\ \midrule \bottomrule
\end{tabular}
\end{table}

To quantify a deviation from GR, we can infer the most probable values of $\delta p_i$ through Bayesian parameter estimation, as discussed in the following subsection.

\subsection{Parameter estimation}
\label{sec:tgr_pe}
Recall our data model $\boldsymbol{d} = \boldsymbol{h} + \boldsymbol{n}$. We denote $\boldsymbol{\theta}(\theta,\delta p_i)$ as the parameter vector generating the signal $\boldsymbol{h}$. It consists of the parameters $\theta$ generating the \texttt{IMRPhenomPv2} waveform and the testing parameters $\delta p_i$ generating the phase deviations from the $\texttt{IMRPhenomPv2}$ waveform.
In practice, the testing parameters are introduced \emph{one at a time}, which is expected to capture a deviation from GR present in multiple phase coefficients while returning narrower credible intervals than introducing multiple coefficients at a time \cite{tgr150914}.

Given the detector output $\boldsymbol{d}$ and prior information $I$, we wish to infer the conditional probability density of $\boldsymbol{\theta}$, referred to as the \emph{posterior}, by invoking Bayes' theorem
\begin{equation}
	P(\boldsymbol{\theta}|\boldsymbol{d},I) = \frac{P(\boldsymbol{d}|\boldsymbol{\theta},I)\times P(\boldsymbol{\theta}|I)}{P(\boldsymbol{d}|I)} \; , \label{bayes_theorem}
\end{equation}
which relates the posterior to three probability densities: the \emph{likelihood} $P(\boldsymbol{d}|\boldsymbol{\theta},I)$, the \emph{prior}  $P(\boldsymbol{\theta}|I)$, and the \emph{evidence}  $P(\boldsymbol{d}|I)$.
During parameter estimation, the evidence, which does not depend explicitly on $\boldsymbol{\theta}$, can be seen as a proportionality constant since $\boldsymbol{d}$ and $I$ are kept fixed. The likelihood and prior are separately discussed below.

Given $\boldsymbol{h}(\boldsymbol{\theta})$, the time series of the output data $\boldsymbol{d}$ uniquely defines a time series of the residual noise $\boldsymbol{d}-\boldsymbol{h}$, which is modeled as Gaussian and stationary. As such, the likelihood is approximated by the Whittle likelihood in  Eq.\ (\ref{whittle}):
\begin{equation}
	P(\boldsymbol{d}|\boldsymbol{\theta},I) \propto \exp\left[- \frac{1}{2}(\boldsymbol{d}-\boldsymbol{h}|\boldsymbol{d}-\boldsymbol{h})\right] \; , \label{logwhittle}
\
\end{equation}
where $(\cdot|\cdot)$ is the \emph{noise-weighted inner product} \cite{cutler}:
\begin{equation}
	(\boldsymbol{a}|\boldsymbol{b}) \equiv \sum_{j=0}^{N /2-1} 4 \operatorname{Re} \left(\frac{\tilde{a}_j^* \tilde{b}_j}{S_{n}(f_j)}\right) \Delta f \; .
\end{equation}
Assuming that noise from multiple detectors, indexed $l$, is uncorrelated, the joint likelihood takes the form
\begin{equation}
	P(\boldsymbol{d}|\boldsymbol{\theta},I) \propto \exp\left[-\frac{1}{2}\sum_l (\boldsymbol{d}_l - \boldsymbol{h}_l  | \boldsymbol{d}_l - \boldsymbol{h}_l)\right] \; . \label{networklikelihood}
\end{equation}

The prior $P(\boldsymbol{\theta}|I)$ incorporates our beliefs about $\boldsymbol{\theta}$ prior to the observation. We follow the default choice of prior in \texttt{LALInference} \cite{veitch2015parameter}, which includes uniform priors for the component masses $m_1$ and  $m_2$, with  $m_2 \leq m_1$, a log-uniform prior for the luminosity distance, an isotropic prior for the sky location of the source and the spin angular momentum vectors of the component black holes, and uniform priors for the remaining parameters.
We note that, in \texttt{LALInference}, the uniform priors specified for component masses are transformed to nonuniform, correlated priors for the chirp mass $\mathcal{M}\equiv (m_1 m_2)^{3 /5} (m_1+m_2)^{-1 /5}$ and the mass ratio $q\equiv m_2 / m_1$ for more efficient sampling \cite{veitch2015parameter}.

In parametrized tests of GR, parameters of primary interest are the testing parameters $\delta p_i$, while the posterior distribution spans the full 16-dimensional parameter space. We therefore compute the \emph{marginalized} posterior distribution for introduced the testing parameter $\delta p_i$:
\begin{equation}
	P(\delta p_i | \boldsymbol{d}, I) =  \int P(\boldsymbol{\theta}|\boldsymbol{d},I) d \theta \; ,
\end{equation}
where $\theta$ denotes the parameters generating the underlying \texttt{IMRPhenomPv2} waveform.

\section{Glitch Mitigation Methods} \label{glitch_mitigation}
In this section, we review four methods that can be applied to mitigate data containing glitches, in which three are used in our investigation, including a frequency-domain filtering method of bandpass filtering, a time-domain filtering method of inpainting, and a glitch model subtraction method using the \texttt{BayesWave} algorithm.

\subsection{Bandpass filtering in frequency domain}\label{maskingfreq}
Assuming stationary and Gaussian noise, components of the noise-weighted inner product from different frequency bins of equal bandwidth and from different detectors contribute linearly to the log likelihood, as seen from Eq.\ (\ref{networklikelihood}). A direct way of removing the glitch in the frequency domain is by excluding the frequency bins containing the glitch from the likelihood calculation.
In \texttt{LALInference}, this can be done by specifying the high-pass and low-pass cutoff frequency for the affected detector such that data containing the glitch are filtered out. Only the passed frequency bins are considered in the likelihood calculation. By default, data are high passed at 20 Hz in \texttt{LALInference} \cite{veitch2015parameter}.

\subsection{Gating and inpainting in time domain}\label{gating}
A similar procedure can be done in the time domain, commonly known as \emph{gating}, in which data containing the glitch are zeroed out by multiplying an inverse window function. The inverse window function reduces the spectral leakage in the frequency domain due to discontinuity of data at the boundary of the region to be zeroed out \cite{harris}. 
 
Gating was adopted in the mitigation of the glitch-overlapped GW170817 signal in LIGO-Livingston during the rapid localization of the source \cite{gw170817}, which successfully led to follow-up electromagnetic observations \cite{gw170817_followup}. However, gating was not used for parameter estimation purposes for the first half of the O3 observing run (O3a) \cite{mitigation_review}. There are concerns over mitigating glitches by gating.
For example, as remarked in Ref.\ \cite{gw170817_mitigation}, gating can introduce errors to parametrized tests of GR, as it affects the signal power in frequency bins that count toward the noise-weighted inner product. 
 
A new method, called \emph{inpainting} or \emph{hole filling} \cite{inpaint}, is developed to address the noise artifacts and statistical bias that may result from gating. After specifying the time interval to be mitigated, new values are assigned for data within the interval, or \emph{hole}, according to an \emph{inpainting filter}, while data outside the hole are unaffected. The inpainting filter depends on the PSD of the stationary Gaussian noise. For inpainted data $\boldsymbol{d}_\text{inp}$, the quantity $(\boldsymbol{d}_\text{inp}|\boldsymbol{h})$ is, by design of the filter, independent of the template waveform  $\boldsymbol{h}$ inside the interval, and $\boldsymbol{d}_\text{inp}$ within the hole is identically zero upon twice whitening by the same PSD \cite{inpaint}. Since the hole can be made arbitrarily narrow, inpainting affects the minimal amount of data if the glitch is localized in time.

Reexpressing the noise-weighted inner product in the likelihood calculation:
\begin{align}
	P(\boldsymbol{d}_\text{inp}|\boldsymbol{h})&\propto \exp\left[-\frac{1}{2} (\boldsymbol{d}_\text{inp}-\boldsymbol{h}|\boldsymbol{d}_\text{inp}-\boldsymbol{h})\right]\nonumber\\ 
						   &=\exp\left[-\frac{1}{2} (\boldsymbol{d}_\text{inp}|\boldsymbol{d}_\text{inp}) + (\boldsymbol{d}_\text{inp}|\boldsymbol{h}) -\frac{1}{2} (\boldsymbol{h}|\boldsymbol{h})\right] \; .
\end{align}

\noindent Given inpainted data $\boldsymbol{d}_\text{inp}$, only the terms  $(\boldsymbol{d}_\text{inp}|\boldsymbol{h})$ and  $(\boldsymbol{h}|\boldsymbol{h})$ differ across waveform templates  $\boldsymbol{h}$; between these two terms, only $(\boldsymbol{d}_\text{inp}|\boldsymbol{h})$ explicitly depends on the inpainted data. As $(\boldsymbol{d}_\text{inp}|\boldsymbol{h})$ is independent of the template waveform inside the hole by design of the filter, inpainted data inside the hole are not expected to contribute to the outcome of parametrized tests.

\subsection{Glitch model subtraction}
\begin{table*}
	\caption{Key specifications of the three mitigation methods.}
	\begin{tabular}{@{}llccc@{}}\toprule\toprule
		Mitigation method\phantom{asdf} & Specification &\phantom{asdf}\textbf{Blip}\phantom{asdf} & \phantom{asdf}\textbf{Tomte}\phantom{asdf} & \phantom{a}\textbf{Scattered light}\phantom{a} \\\toprule
		\textbf{Bandpass} & High-pass cutoff (Hz) & 20 & 105 & 40 \\
		 & Low-pass cutoff (Hz) & 60 & 511.875 & 511.875\\\midrule
	\textbf{Inpainting} & Hole duration (s) & 0.005 & 0.040 & $\dots^*$ \\
			   & Sampling rate (Hz) & 4096 & 4096 & $\dots^*$ \\\midrule
	\textbf{Glitch model} & Segment length (s) & 4 & 4 & 8\\
	\textbf{subtraction}    &  High-pass cutoff (Hz) & 20 & 20 & 8\\
			    & Sampling rate (Hz) & 2048 & 2048 & 2048 \\
			    & $Q_\mathrm{max}$ & 40 & 40 & 200\\
			    & $D_{\mathrm{max}}$ & 100 & 100 & 200 \\\midrule\toprule\\[-2.1ex]
			    \multicolumn{5}{l}{\scriptsize $^*$Inpainting is replaced by discarding data from the detector which the scattered-light glitch is present.}
	\end{tabular}
	
	\label{table:specifications}
\end{table*}
\begin{figure*}
	\includegraphics[height=0.20\textheight]{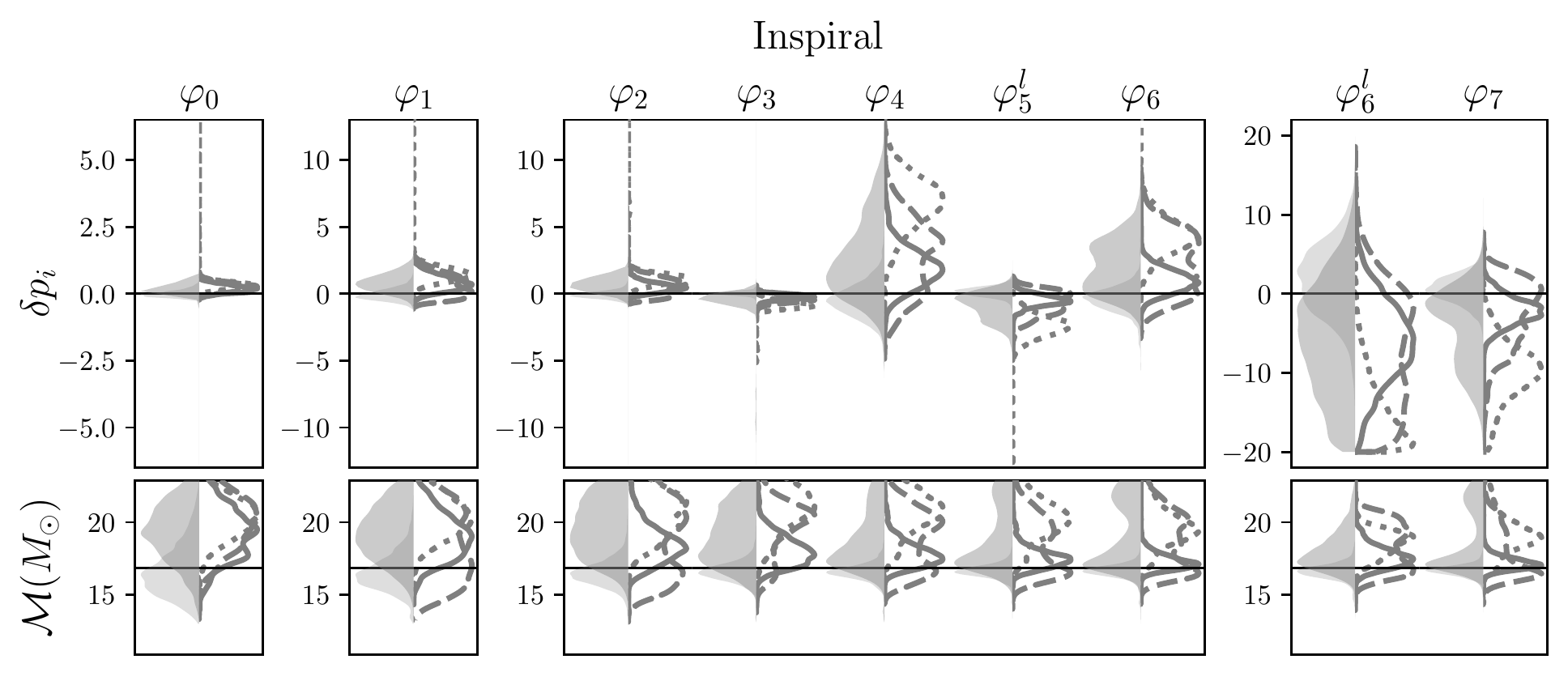}
	\includegraphics[height=0.20\textheight]{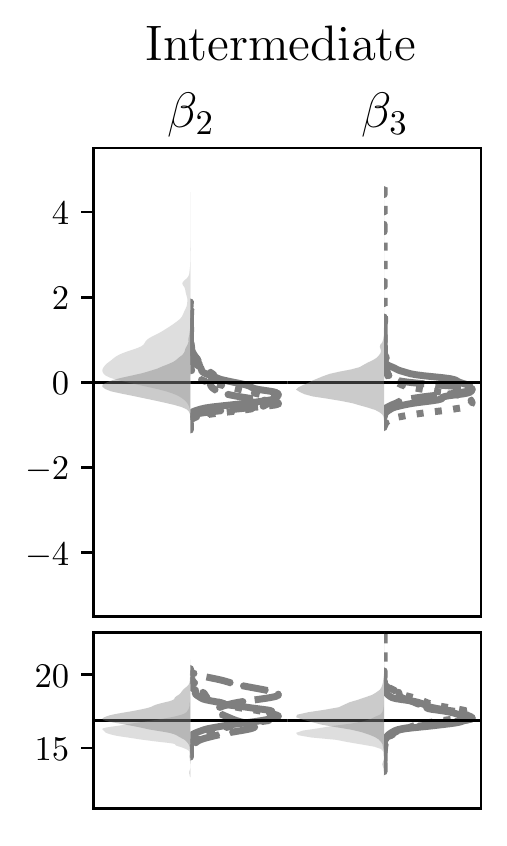}
\includegraphics[height=0.20\textheight]{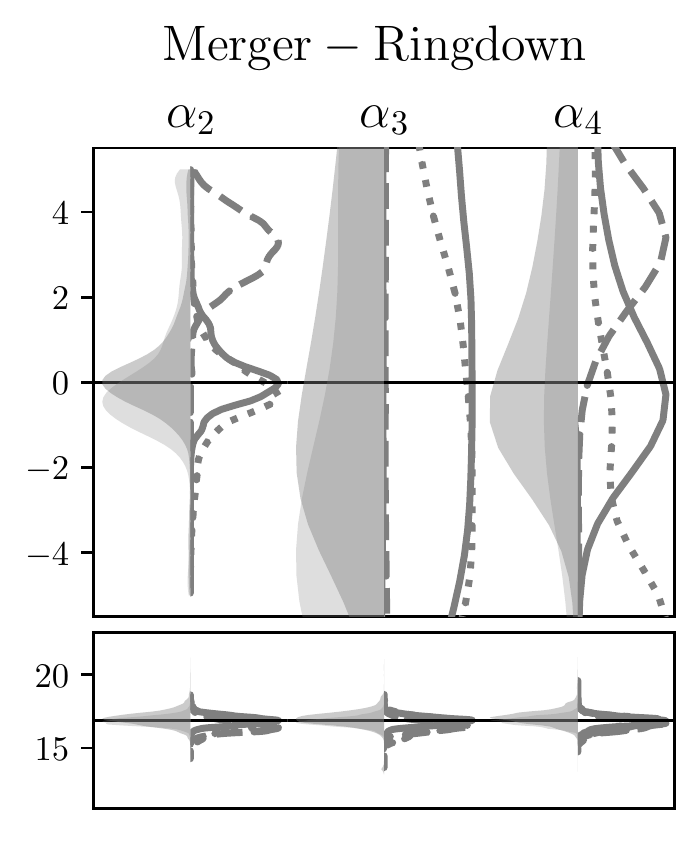}\\
\caption{Posterior distributions of testing parameters (top) and recovered chirp mass (bottom) obtained by performing parametrized tests of GR on five data realizations of a simulated GW190828\_065509-like signal in stationary Gaussian noise. Each shading and line style (left, light-gray shade and dark-gray shade; right, solid, dotted and dashed line) represents one data realization. The simulated noise is colored by the representative best LIGO-Hanford, LIGO-Livingston and Virgo detector PSD estimates during O3a.}
	\label{fig:quantile}
\end{figure*}
The \texttt{BayesWave} \cite{bayeswave, bayeswave2} algorithm models the GW signal and glitches in each detector using a variable number of wavelets, such as sine-Gaussian wavelets. Mitigated, or \emph{deglitched}, data are generated by subtracting the glitch model from the original data. Using Bayesian inference, the output data in each detector are modeled as a superposition of a GW signal $\boldsymbol{h}$, stationary Gaussian noise $\boldsymbol{n}_G$ and glitches $\boldsymbol{g}$:
\begin{align}
	\boldsymbol{d} = \boldsymbol{h} + \boldsymbol{n}_G + \boldsymbol{g} \; .
\end{align}
While both the GW signal and glitches are nonstationary and non-Gaussian, coherent features across data from multiple detectors are modeled by the signal model and independent features are modeled by the glitch model \cite{bayeswave_new}. A transdimensional reversible jump Markov chain Monte Carlo algorithm is used to sample models with different number of wavelets or with wavelets of different parameter values. 
The most probable model is inferred through Bayesian inference by comparing the evidence $P(\boldsymbol{d}|M_i,I)$ for different models  $M_i$: Given data $\boldsymbol{d}$ and prior information  $I$, we define the \emph{odds} $O^1_2$ between two competing models $M_1$ and  $M_2$ as
\begin{align}
	O^1_2 \equiv \frac{P(M_1|\boldsymbol{d},I)}{P(M_2|\boldsymbol{d},I)} = \frac{P(M_1|I)}{P(M_2|I)}\times \frac{P(\boldsymbol{d}|M_1,I)}{P(\boldsymbol{d}|M_2,I)} \; , \label{eq:model_selection}
\end{align}
where the equality on the right is obtained by invoking Bayes' theorem. The model $M_1$ will be more probable than model $M_2$ if the odds $O^1_2$ are larger than 1. To express our ignorance toward the probability of models prior to observation, we can set the first term on the rightmost in Eq.\ (\ref{eq:model_selection}), called the \emph{prior odds}, to unity. The odds can then be obtained by comparing the evidences of the two models. In \texttt{BayesWave}, the evidences are calculated through thermodynamic integration \cite{bayeswave}. Once the most probable glitch+signal model is inferred, the glitch model is subtracted from the data.

The \texttt{BayesWave} algorithm was first used to remove the glitch which overlapped with the GW170817 signal during parameter estimation \cite{gw170817} and was regularly used to mitigate glitch-overlapped signals during O3a \cite{gwtc2}. Reference \cite{gw170817_mitigation} concluded that parameter recovery results using data reconstructed by \texttt{BayesWave} are unbiased.
In the context of tests of GR, which are designed to detect small deviations from GR waveforms, the subtraction of sine-Gaussian wavelets by \texttt{BayesWave} may alter the GW signal to an extent which may be reported as a false violation of GR. This is not observed in our results.
\section{Results of Glitches overlapping a GW190828\_065509-like Signal}\label{sec:results}
We are motivated to consider a signal similar to that of the high-mass-ratio BBH-merger event GW190828\_065509 \cite{gwtc2}, in which the mitigation of potential glitches overlapping the event in L1 through bandpass filtering resulted in pathological features in parametrized tests of GR \footnote{Private communication with Rico K.\ L.\ Lo.}.
Values of some selected generating parameters of the GW190828\_065509-like signal are tabulated in Table \ref{tab:generating_parameters}.
\begin{table}[ht]
	\caption{Injected values of some selected generating parameters of a GW190828\_065509-like signal using the \texttt{IMRPhenomPv2} waveform model. The GW190828\_065509-like signal is taken to be the \emph{maximum likelihood} waveform inferred for real GW190828\_065509 data using the \texttt{IMRPhenomPv2} template waveform model. Despite the large injected values for the component spins, the inferred posterior distributions of the component spins are flat throughout the prior range for the GW190828\_065509 and simulated GW190828\_065509-like signals.}
	\centering
	\begin{tabular}[t]{lc}
		\toprule\toprule
		Waveform parameter&Value\\
		\toprule
		Chirp mass $\mathcal{M}$ ($M_\odot$)&16.86\\
		Mass ratio $q$ &0.14\\
		Dimensionless primary spin magnitude $a_1$ &0.92\\
		Dimensionless secondary spin magnitude $a_2$ &0.75\\
		Right ascension $\alpha$ (rad) & 2.54\\
		Declination $\delta$ (rad) &$-0.84$\\
		Luminosity distance $D_L$ (Mpc) & 1021\\
		\toprule\midrule
	\end{tabular}
	\label{tab:generating_parameters}
\end{table}

We first present the expected results of parametrized tests of GR in the absence of glitches by coherently injecting the simulated GW190828\_065509-like signal, generated with an \texttt{IMRPhenomPv2} waveform model, into five realizations of simulated stationary, Gaussian noise colored with the representative best (cleaned) PSD of the LIGO-Hanford (H1), LIGO-Livingston (L1) and Virgo (V1) detectors during O3a.
The posterior distributions of the testing parameters and the recovered chirp mass are plotted in Fig.\ \ref{fig:quantile}.
\begin{figure*}
	\subfloat[$\delta\varphi_3$]{\includegraphics[width=0.32\textwidth]{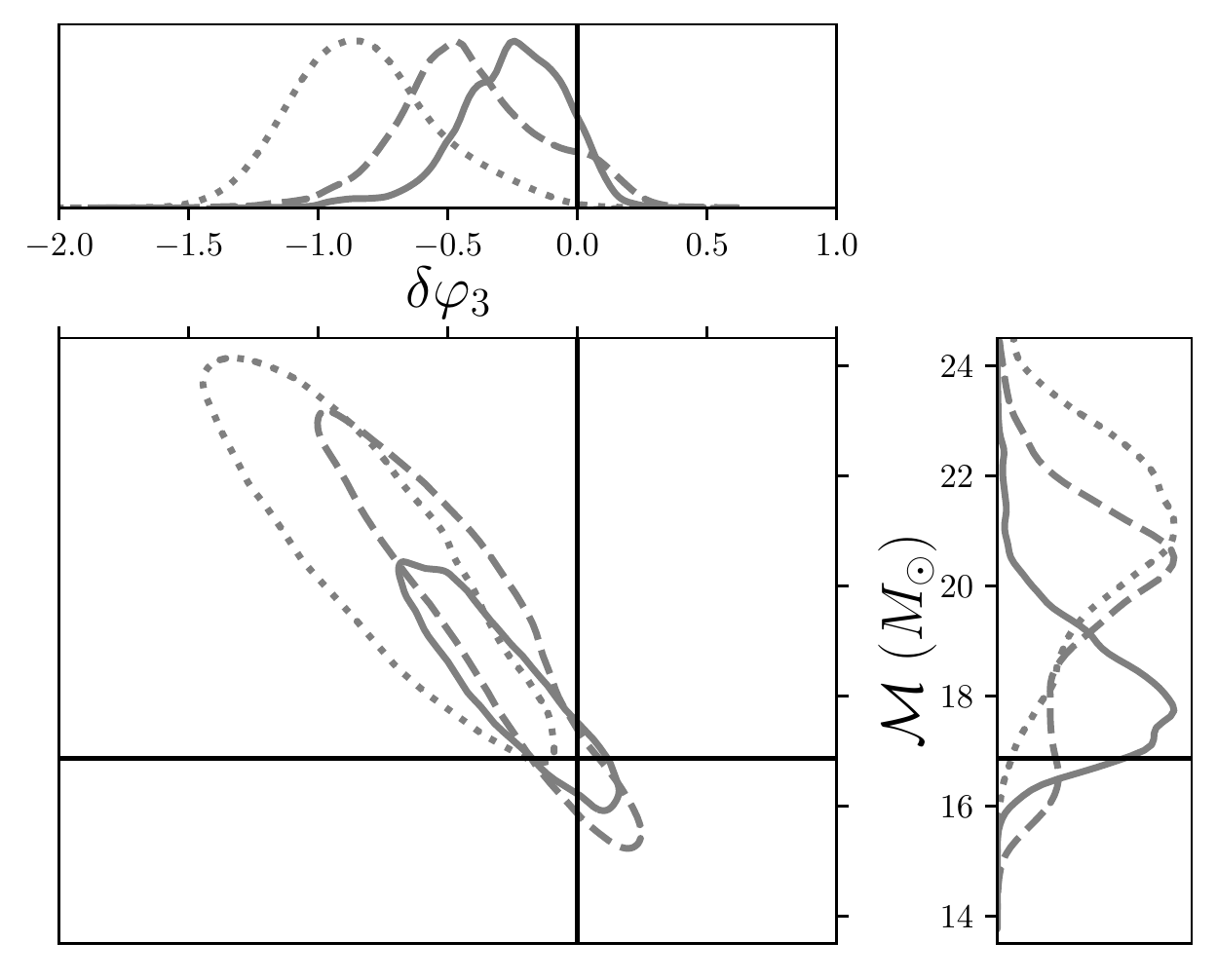}\label{fig:gaussian_corner_dchi3}}\hspace{0.01\textwidth}
	\subfloat[$\delta\varphi_4$]{\includegraphics[width=0.32\textwidth]{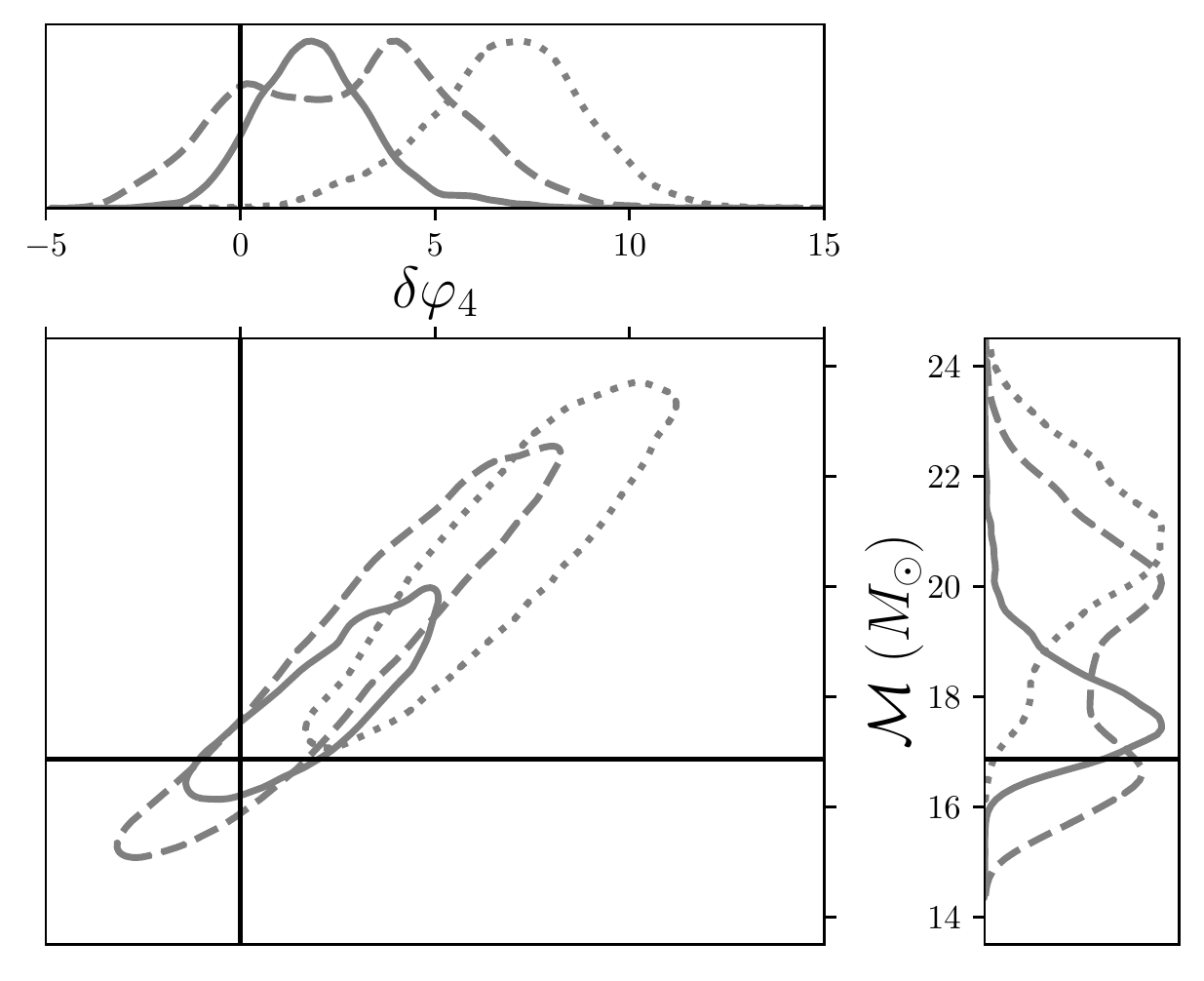}\label{fig:gaussian_corner_dchi4}}\hspace{0.01\textwidth}
	\subfloat[$\delta\alpha_2$]{\includegraphics[width=0.32\textwidth]{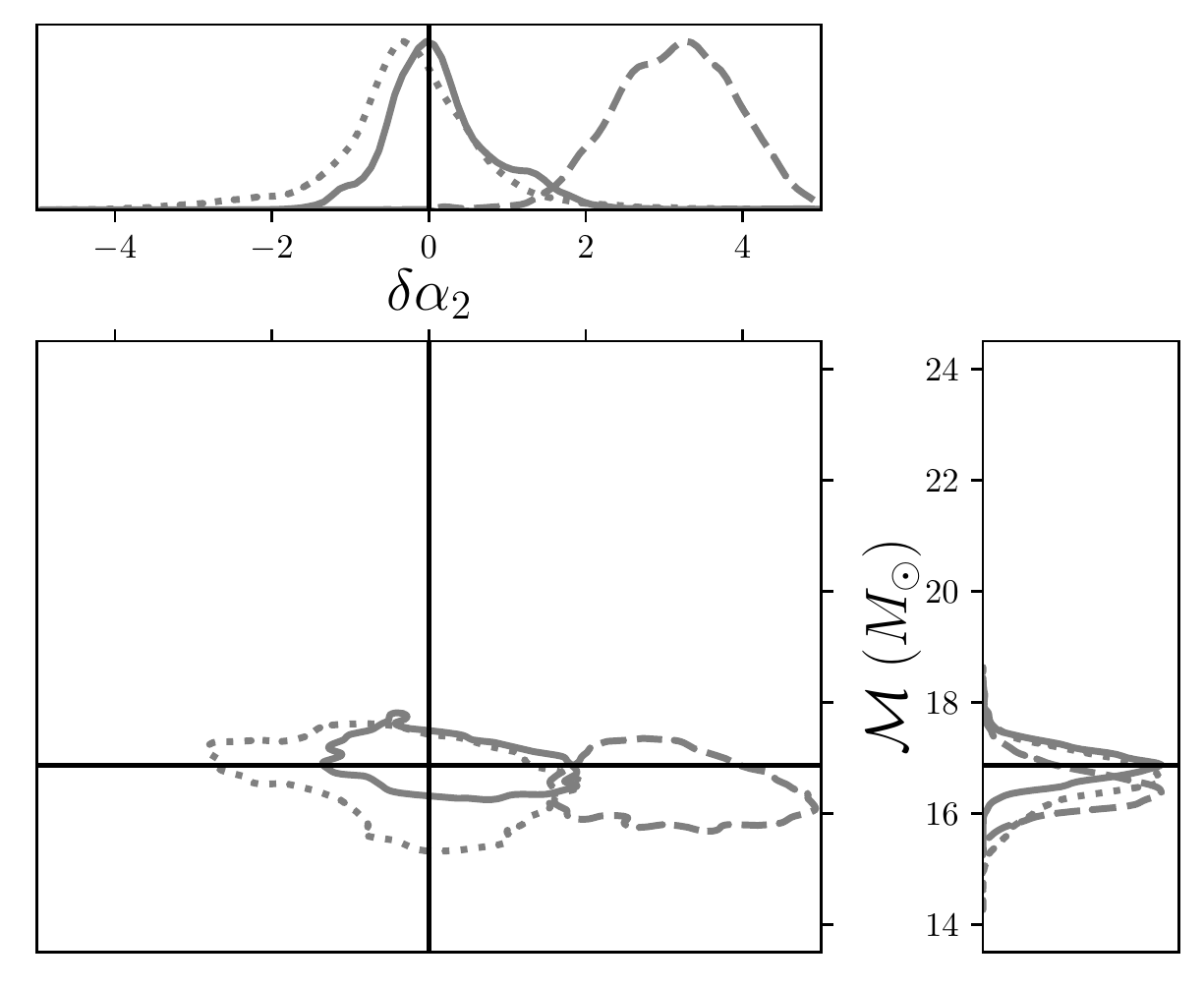}\label{fig:gaussian_corner_dalpha2}}
	\caption{Marginalized distributions of testing parameters (left, $\delta\varphi_3$; middle, $\delta\varphi_4$; right,  $\delta\alpha_2$) and chirp mass. The contours of the two-dimensional distributions show 90\% credible regions. Each line style represents the data realization of the simulated GW190828\_065509-like signal in stationary Gaussian noise with the corresponding line style in Fig.\ \ref{fig:quantile}. The vertical and horizontal black lines denote the GR value of the testing parameters and the injected value of chirp mass, respectively.}
	\label{fig:gaussian_corner}
\end{figure*}

Beneath the posteriors of testing parameters, the posteriors of recovered chirp mass are also plotted as an indicator of the sampling performance. With an extra degree of freedom introduced by inspiral PN testing parameters, broad and occasional multimodal distributions can be observed due to strong correlations between the inspiral PN testing parameters and chirp mass, as demonstrated in Figs.\ \ref{fig:gaussian_corner_dchi3} and \ref{fig:gaussian_corner_dchi4}.
Such degeneracies could also bring the one-dimensional marginalized distributions of the inspiral PN testing parameters away from zero, e.g., for the dotted distributions in Figs.\ \ref{fig:gaussian_corner_dchi3} and \ref{fig:gaussian_corner_dchi4}, where the GR value of 0 is excluded at 90\% credibility. 
These exclusions of 0 should not be counted toward a violation of GR, as the bias introduced by sampling can be clearly identified.

Correlations between inspiral PN testing parameters and mass parameters can be expected from theoretical grounds, as $\delta\varphi_j$ contain the term proportional to $\eta^{-1}M^{(j-5)/3}$, where $\eta$ and $M$ are the symmetric mass ratio and total mass, respectively, with $\delta\varphi_0$ being completely degenerate with chirp mass. The degeneracy of inspiral PN testing parameters with the chirp mass used during sampling may be amplified by the weakness of the signal, as we are injecting a weak signal \footnote{The event GW1902828\_065509 has a full-signal optimal SNR of 9.9 - weakest of all events considered for parameterized tests of GR in GWTC-1 and GWTC-2 combined \cite{tgrgwtc, o3a_tgr}.}.

We, however, note that the exclusion of 0 in simulated Gaussian noise of the phenomenological merger-ringdown parameter $\delta\alpha_2$ for one considered data realization cannot be explained in a similar way, as a degeneracy with chirp mass cannot be observed in Fig.\ \ref{fig:gaussian_corner_dalpha2}. The source of this bias is not identified in the present study, but it may be due to the weakness of the injected signal, as similar anomalies arise when substantial signal power is discarded in Fig.\ \ref{fig:highpass}.

The same signal is then injected into real data from three detectors (H1, L1, and V1)
at times when all three detectors are operating in the science mode and with glitches present in either H1 or L1
\footnote{Calibrated, cleaned data from H1 and L1 are taken from the strain channel \texttt{DCS-CALIB\_STRAIN\_CLEAN\_C01}.  Reproduced data from V1 are taken from the strain channel \texttt{Hrec\_hoft\_V1O3Repro1A\_16384Hz} \cite{gwosc}}.
Glitches from the \emph{blip}, \emph{tomte}, and \emph{scattered-light} classes are chosen, as these classes of glitches have the highest occurrence rates in O3a \cite{glitcho3a}. The glitches used in our study are further chosen so that their duration and peak frequency are representative of their corresponding glitch classes. The chosen blip, tomte, and scattered-light glitch are present at GPS time around 1253103382, 1252901859, and 1253416025 at H1, L1, and H1, respectively.

The GW190828\_065509-like signal is injected coherently into the three detectors such that each glitch overlaps with the signal at the inspiral, intermediate, and merger-ringdown stage in the time domain for different data samples.
The three stages in the time domain are defined as the time intervals when the instantaneous frequencies of the signal are in the corresponding three stages in the frequency domain discussed in Sec.\ \ref{waveform_model}, respectively.
The boundaries of the three stages of the signal in the time and frequency domain are marked in the $Q$ scans by vertical and horizontal white lines, respectively.

After preparing the data samples, we applied the glitch mitigation methods of bandpass filtering, inpainting, and \texttt{BayesWave} glitch subtraction as described in Sec.\ \ref{glitch_mitigation} on detector data in which glitches are present. We then performed parametrized tests of GR on the unmitigated and mitigated samples. The specifications of the three glitch mitigation methods are tabulated in Table \ref{table:specifications}. We referred to Ref.\ \cite{bayeswave_new2} for \texttt{BayesWave} specifications, whereas bandpass cutoff frequencies and inpainting hole duration are chosen to exclude time-frequency bins affected by the glitch.

\subsection{Blip glitch} \label{sec:results_blip}
\begin{figure*}
	\includegraphics[width=\textwidth]{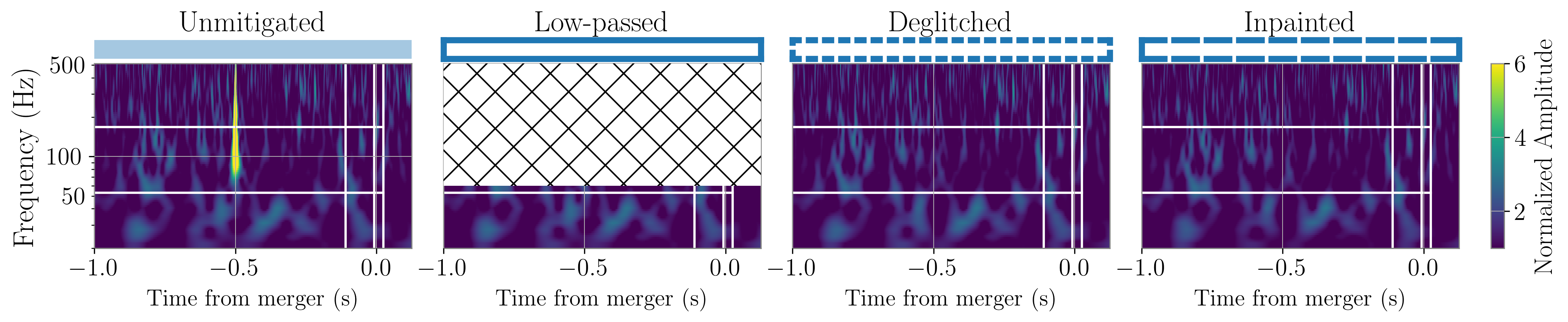}\\[-1.0ex]
	\subfloat[Simulated GW190828\_065509-like signal overlapped with a H1 blip glitch at inspiral stage in the time domain.]{\includegraphics[height=0.15\textheight]{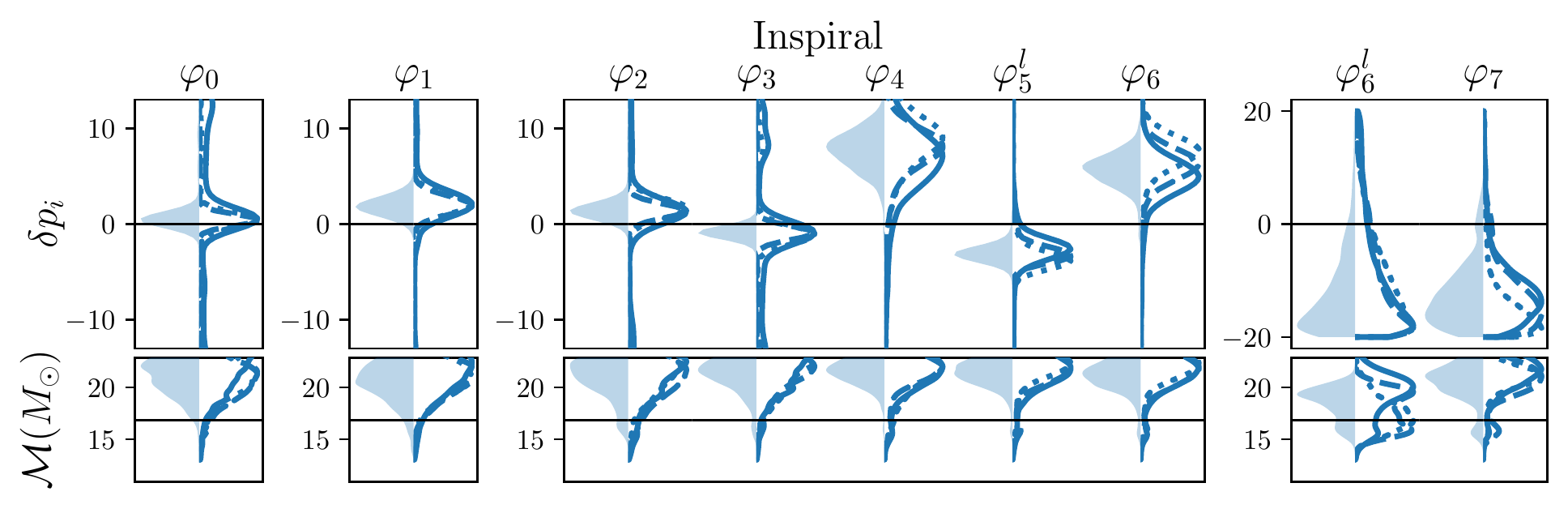}\label{fig:blip_insp}
	\includegraphics[height=0.15\textheight]{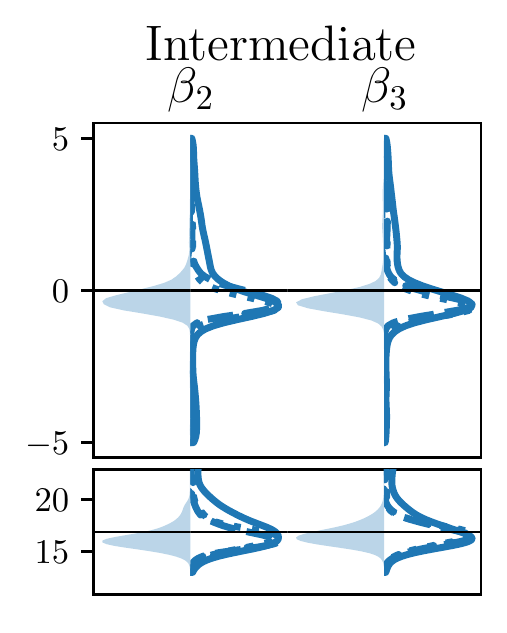}
\includegraphics[height=0.15\textheight]{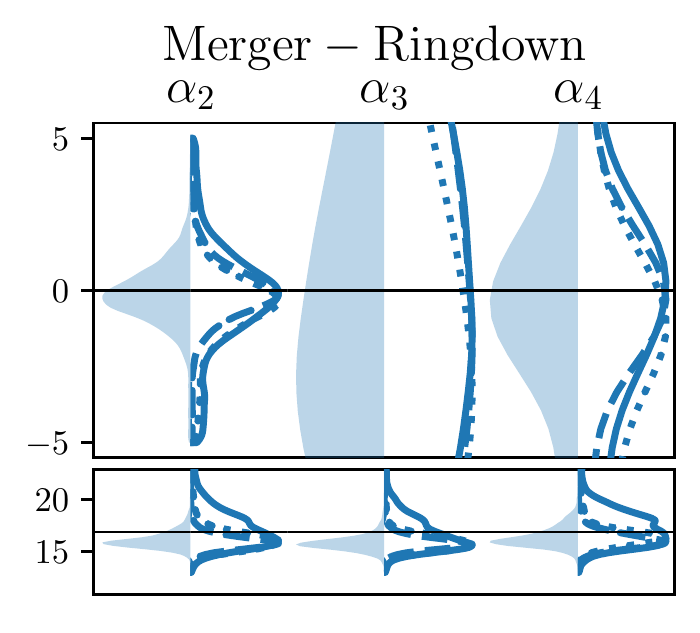}}\\
	\includegraphics[width=\textwidth]{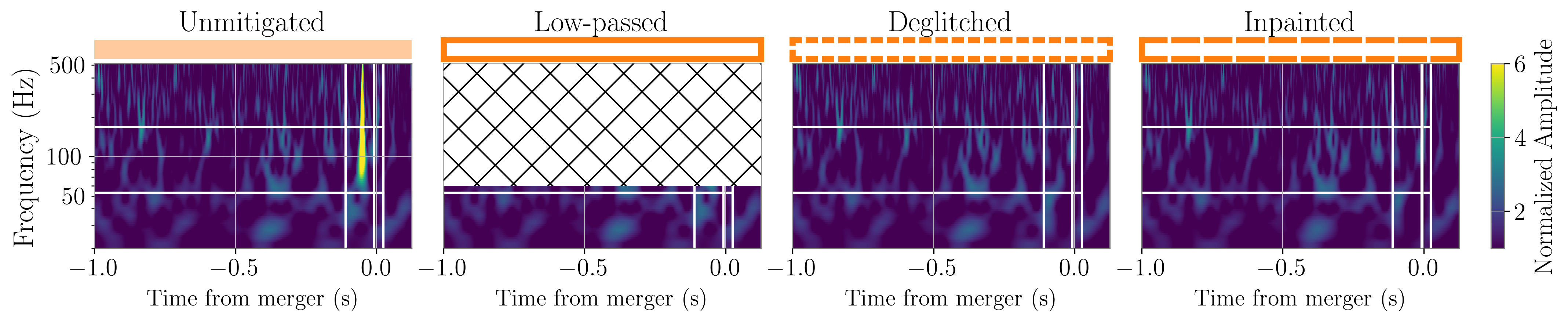}\\[-1.0ex]
	\subfloat[Simulated GW190828\_065509-like signal overlapped with a H1 blip glitch at intermediate stage in the time domain.]{\includegraphics[height=0.15\textheight]{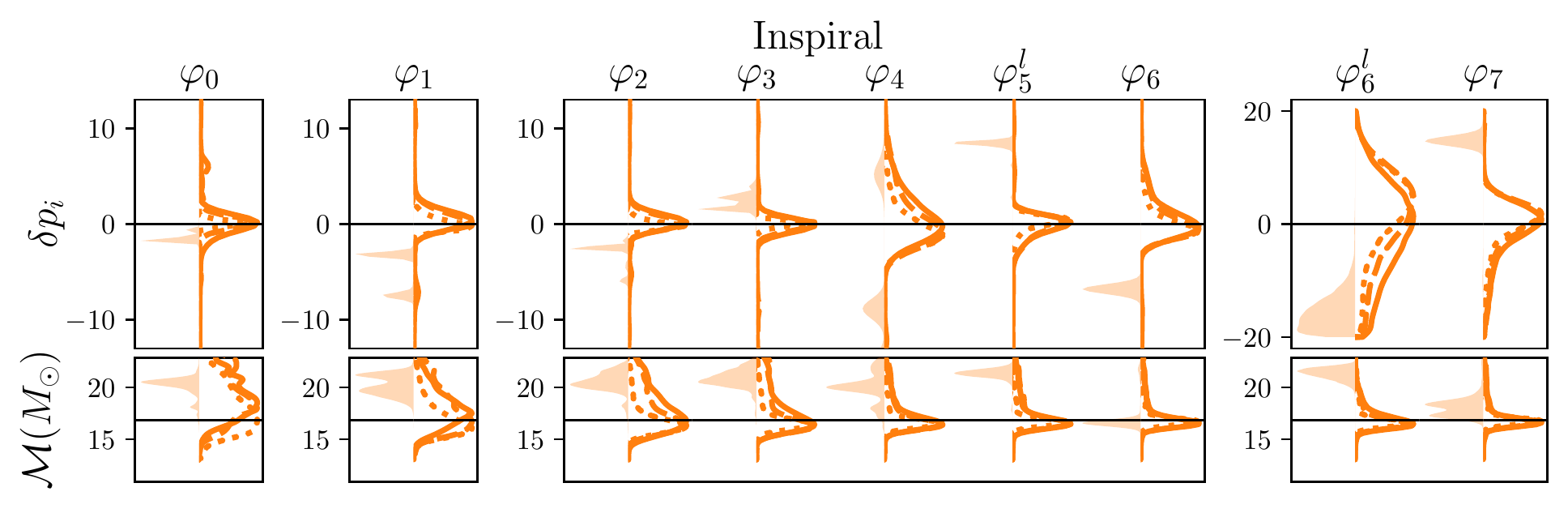}\label{fig:blip_inter}
	\includegraphics[height=0.15\textheight]{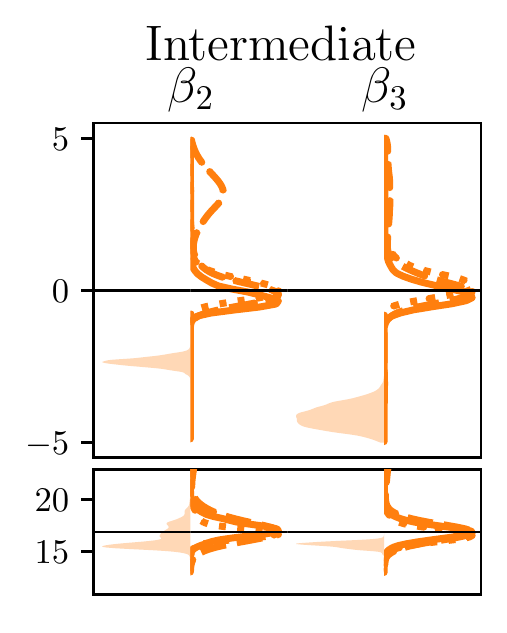}
\includegraphics[height=0.15\textheight]{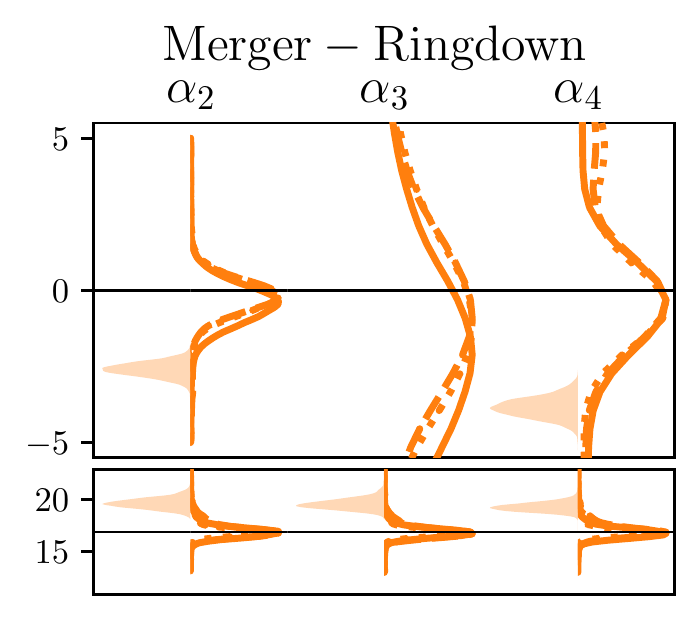}}\\
	\includegraphics[width=\textwidth]{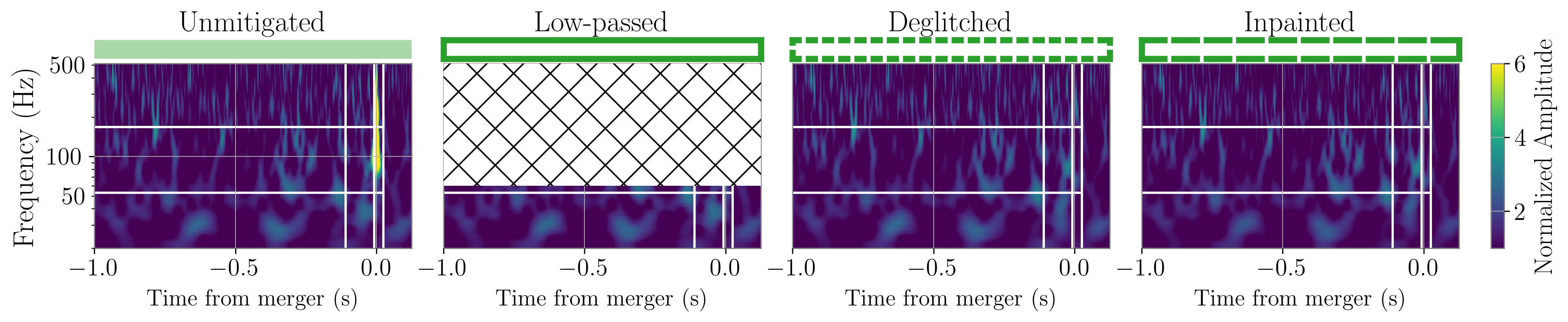}\\[-1.0ex]
	\subfloat[Simulated GW190828\_065509-like signal overlapped with a H1 blip glitch at merger-ringdown stage in the time domain.]{\includegraphics[height=0.15\textheight]{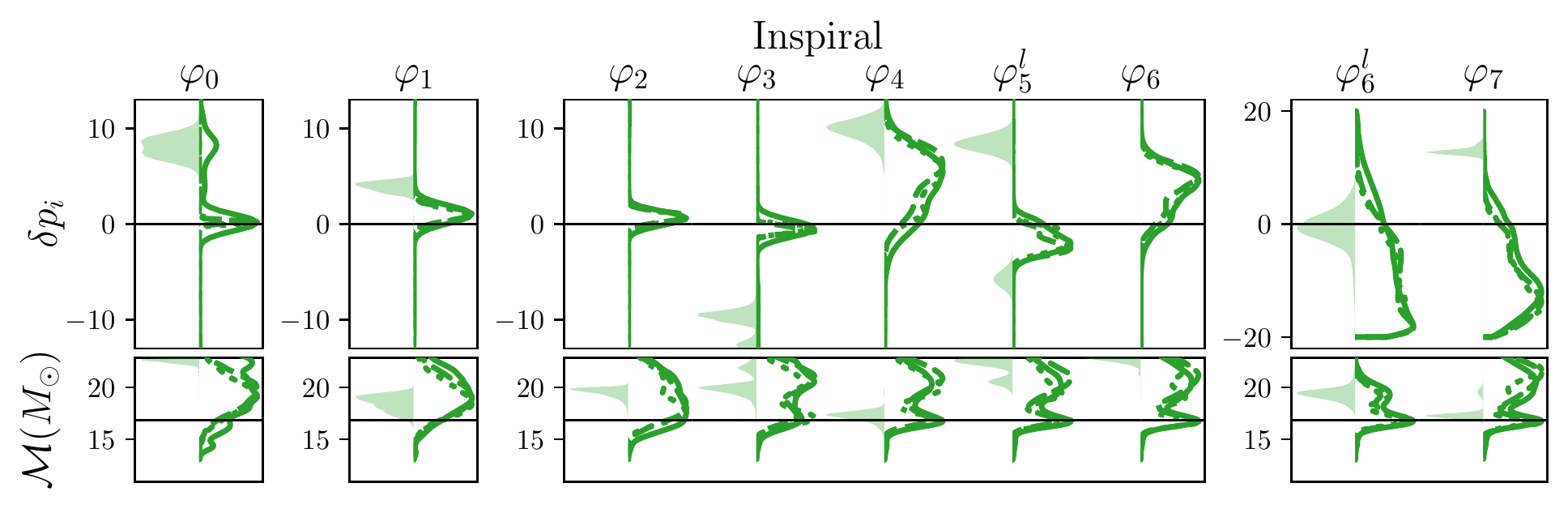}\label{fig:blip_mr}
	\includegraphics[height=0.15\textheight]{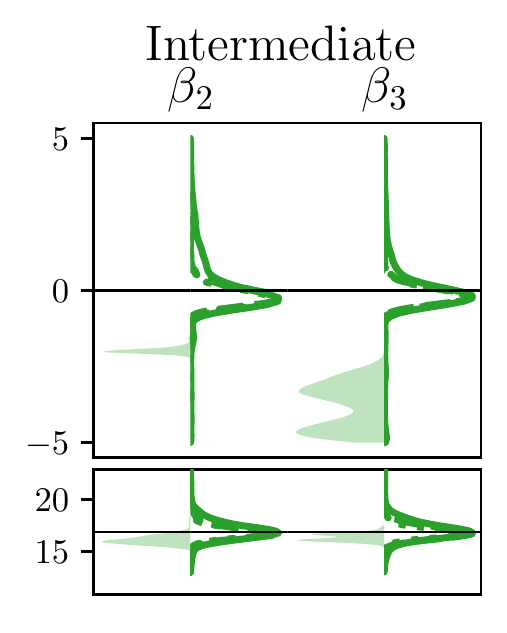}
\includegraphics[height=0.15\textheight]{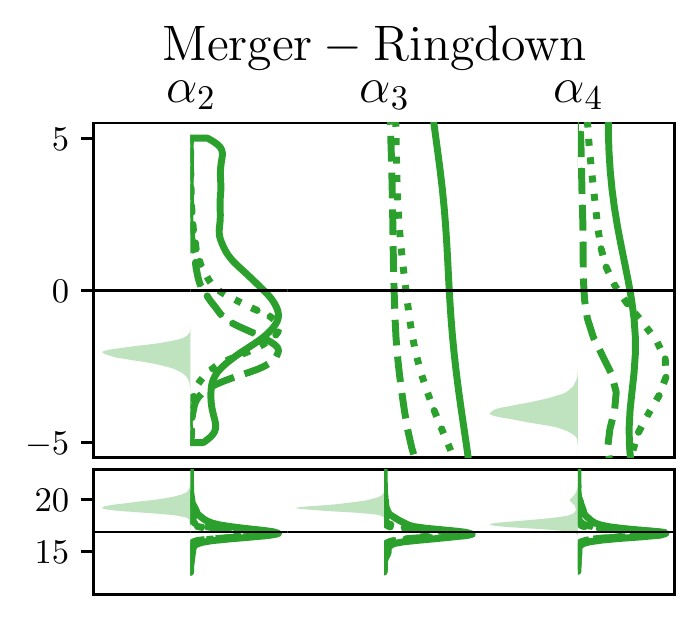}}\\
\caption{Top subfigures: $Q$ scans of the unmitigated and mitigated data samples. Vertical and horizontal white lines denote the boundaries of different stages of coalescence in the time and frequency domain, respectively. Bottom subfigures: Posterior distributions of testing parameters (top) and the recovered chirp mass (bottom) obtained by performing parametrized tests on unmitigated (left of violin plot) blip-glitch-overlapped signals during a three-detector observation. The corresponding mitigated cases (right of violin plot) with bandpass filtering (solid line), \texttt{BayesWave} glitch model subtraction, also called \emph{deglitching} (dotted line), and inpainting (dashed line) are also plotted. The GR value of the testing parameters and the injected value of chirp mass are indicated by vertical black lines.}
	\label{fig:blip_violins}
\end{figure*}
 Blip glitches are short-duration, broadband glitches characterized by their teardrop shape as seen in time-frequency representations. A $Q$ scan of a blip glitch is shown on the top left in Fig.\ \ref{fig:glitch} \cite{glitch2}. The sources and coupling of blip glitches are not well understood \cite{gwtc2}.

The simulated GW190828\_065509-like signal is coherently injected into H1, L1, and V1 in a way that a blip glitch at H1 overlaps with the signal at the inspiral, intermediate and merger-ringdown stages in the time domain.
Mitigation methods are applied to H1 data, and parametrized tests of GR are performed on the unmitigated and mitigated data. The posteriors of the testing parameters are plotted on the left and right side of each violin plot in Fig.\ \ref{fig:blip_violins}, respectively, while the unmitigated and mitigated data from H1 are represented by $Q$ scans.

The stages of coalescence where violations of GR are observed show no correlation with those overlapped by the glitch in the time or frequency domain:
Violations of GR can be observed for testing parameters from all stages of coalescence when the blip glitch overlaps with the signal in the intermediate or merger-ringdown stage in the time domain, even though the blip glitch contributes excess power only to intermediate and merger-ringdown frequency bands.
 
\begin{figure}
	\includegraphics[width=0.85\linewidth]{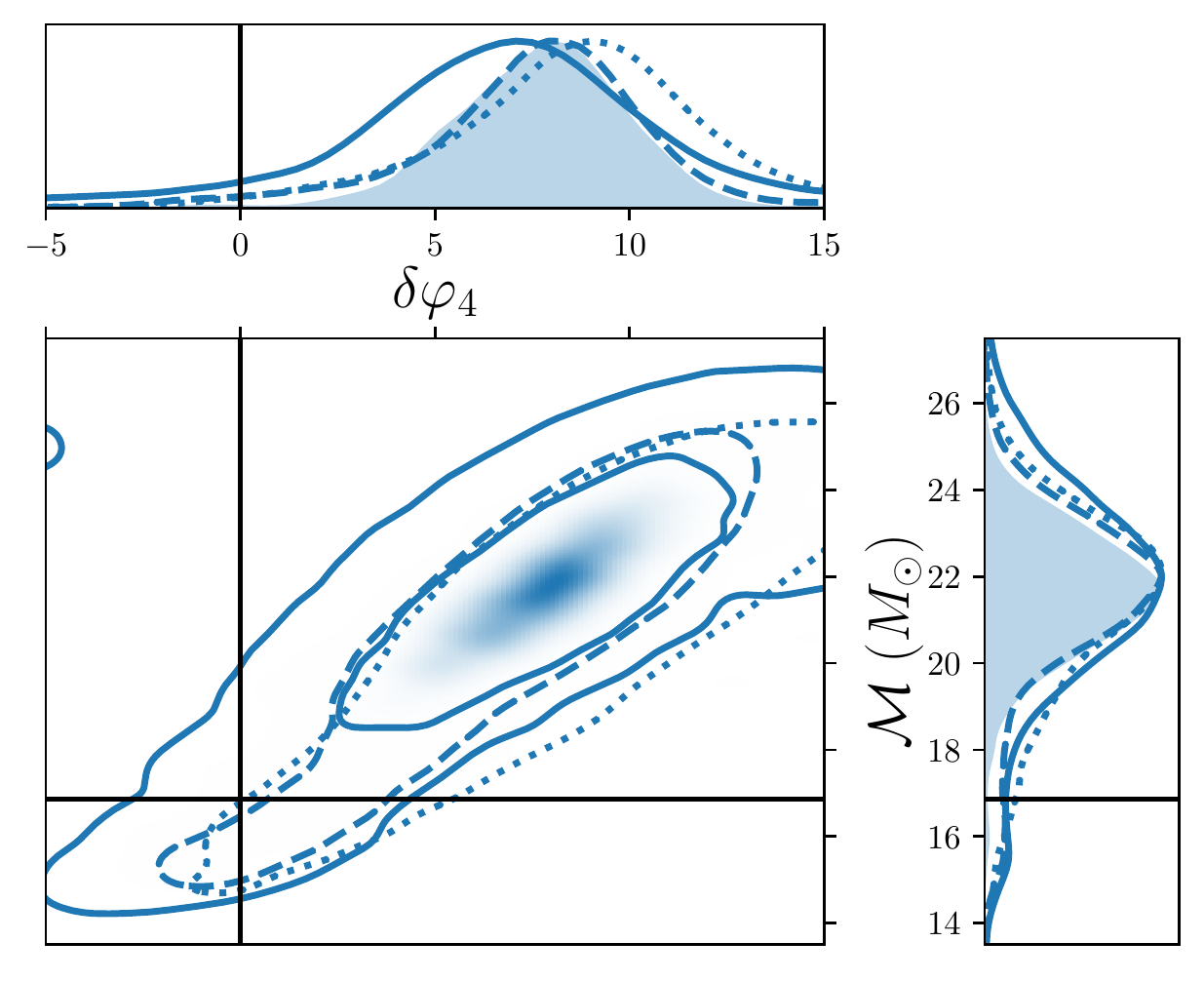}\\
	\caption{Marginalized distributions of $\delta\varphi_4$ and chirp mass. The contours of the two-dimensional distributions show 90\% credible regions. The distributions are obtained by performing parametrized tests on unmitigated (shaded), bandpassed (solid line), glitch-subtracted (dotted line), and inpainted (dashed line) blip-glitch-overlapped signal as shown in Fig.\ \ref{fig:blip_insp}. The vertical and horizontal black lines denote the GR value of the testing parameter and the injected value of chirp mass, respectively.}
	\label{fig:blip_insp_corner}
\end{figure}

No observable effects on parametrized tests of GR are observed when the blip glitch temporally overlaps the signal at the inspiral stage, suggested by the matching posterior distributions without and with the glitch removed through independent methods of low passing to 60 Hz, deglitching, and inpainting, though the GR value of 0 is located only at the far tails of distributions for the inspiral PN testing parameters due to the degeneracy with chirp mass, as demonstrated in Fig.\ \ref{fig:blip_insp_corner}.

Comparing the unmitigated and mitigated results of the glitch overlapping the intermediate and merger-ringdown stage in the time domain, all three mitigation methods of low passing, inpainting, and deglitching can reduce false violations of GR by bringing posteriors of testing parameters from distributions that exclude the GR value of 0 at 90\% credibility to one that peaks close to 0 [e.g., $\delta\varphi_3, \delta\varphi_4, \delta\beta_2,\delta\beta_3$, and $\delta\alpha_2$ in Fig.\ \ref{fig:blip_mr}]. The posterior distributions of the testing parameters for mitigated samples match each other closely, indicating that the mitigation methods did not contribute extra effects on parametrized tests of GR in these three cases.
 Significant improvements in parametrized tests of GR after removal of the blip glitch suggest that false violations are attributed to the presence of the glitch.

Upon introduction of an extra degree of freedom by the parametrized deviations, the blip glitch leads to bias in other intrinsic variables such as the chirp mass. This is reflected by the sharp peaks in the chirp mass posteriors which exclude the injected value at 90\% credibility [e.g., $\delta\alpha_2$, $\delta\alpha_3$, and $\delta\alpha_4$ in Fig.\ \ref{fig:blip_inter} and  $\delta\varphi_2$, $\delta\varphi_{5l}$, and $\delta\varphi_6$ in Fig.\ \ref{fig:blip_mr}]. The independent glitch mitigation methods bring the peak values close to the injected value, although multimodal and broad distributions for chirp mass can still be noted after mitigation due to its degeneracy with inspiral PN testing parameters.

We quote representative figures of the network matched-filter signal-to-noise ratio (henceforth SNR) for unmitigated and mitigated samples, recovered with \texttt{IMRPhenomPv2} (without introducing parametrized deviations) using data samples in Fig.\ \ref{fig:blip_mr}, where the simulated signal overlaps with the glitch at the merger-ringdown stage in the time domain.
The SNR are $20.76^{+0.18}_{-0.28}$ before mitigation, $8.26^{+0.41}_{-1.07}$ after low passing, $9.67^{+0.27}_{-0.53}$ after glitch subtraction, and $9.93^{+0.26}_{-0.47}$ after inpainting. Symmetric 90\% credible intervals are denoted by the subscripts and superscripts. 
\subsection{Tomte glitch} \label{sec:results_tomte}
\begin{figure}
	\includegraphics[width=0.85\linewidth]{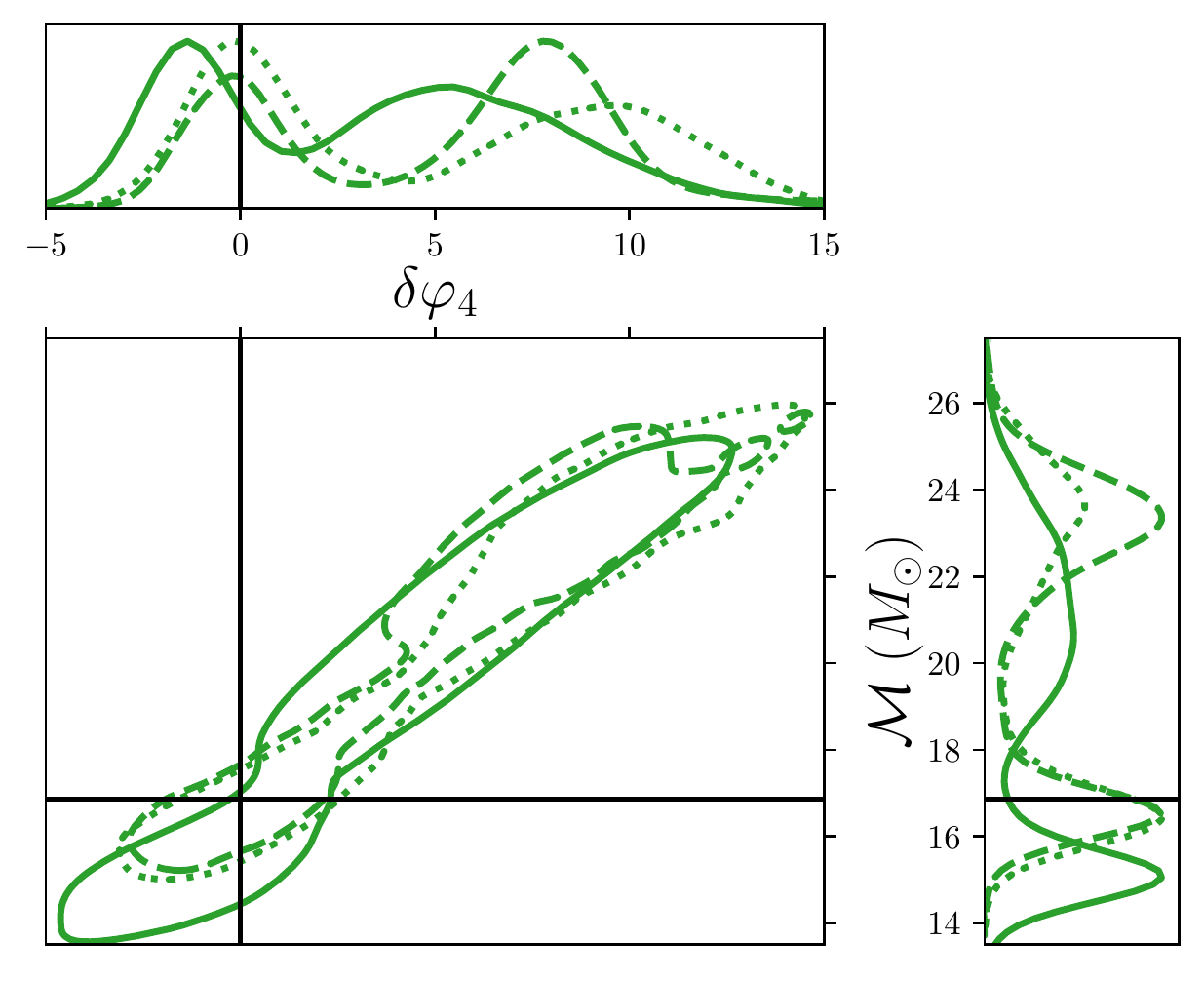}
	\caption{The same as Fig.\ \ref{fig:blip_insp_corner}, but for the tomte-glitch-overlapped signal as shown in Fig.\ \ref{fig:tomte_mr}. The unmitigated distributions are out of the boundaries of the figure.}
	\label{fig:tomte_mr_corner}
\end{figure}
\begin{figure*}
	\includegraphics[width=\textwidth]{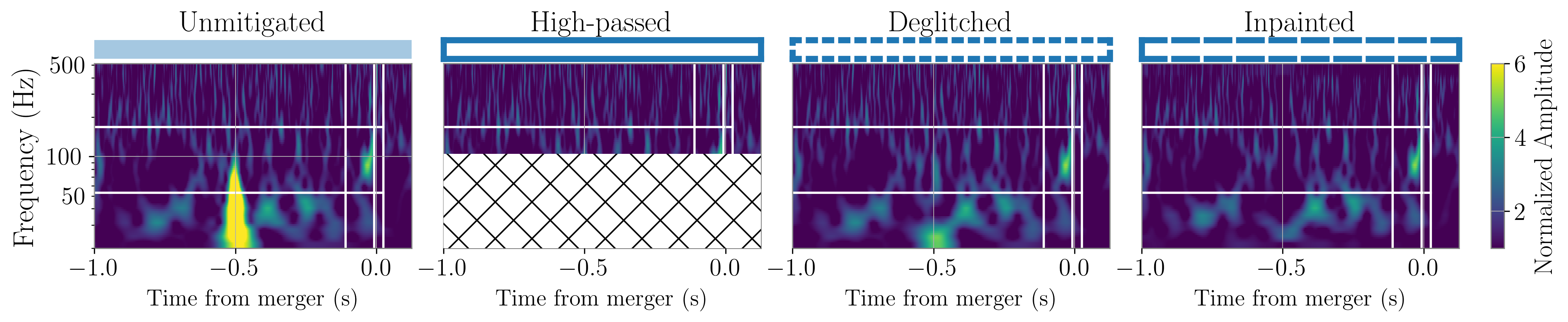}\\[-1.0ex]
	\subfloat[Simulated GW190828\_065509-like signal overlapped with a L1 tomte glitch at inspiral stage in the time domain.]{\includegraphics[height=0.15\textheight]{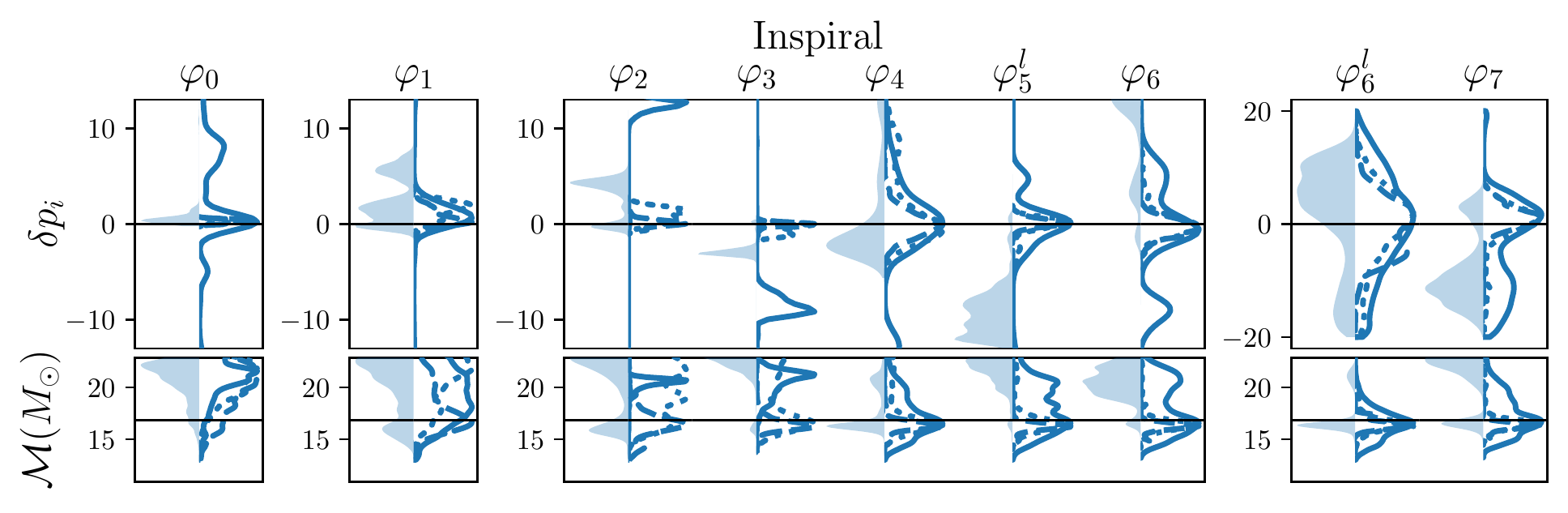}\label{fig:tomte_inspiral}
	\includegraphics[height=0.15\textheight]{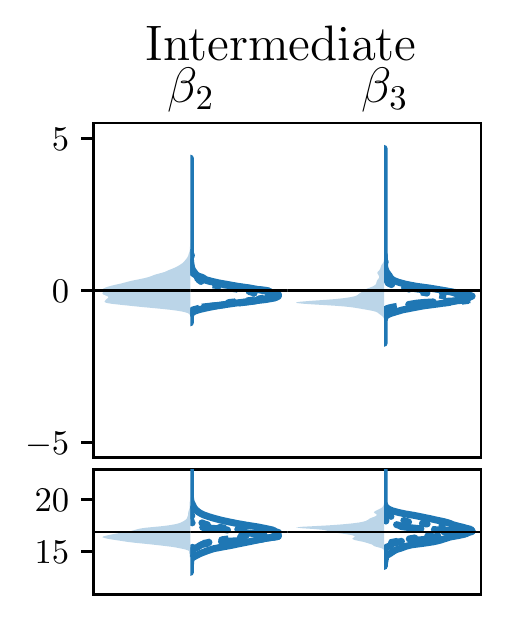}
\includegraphics[height=0.15\textheight]{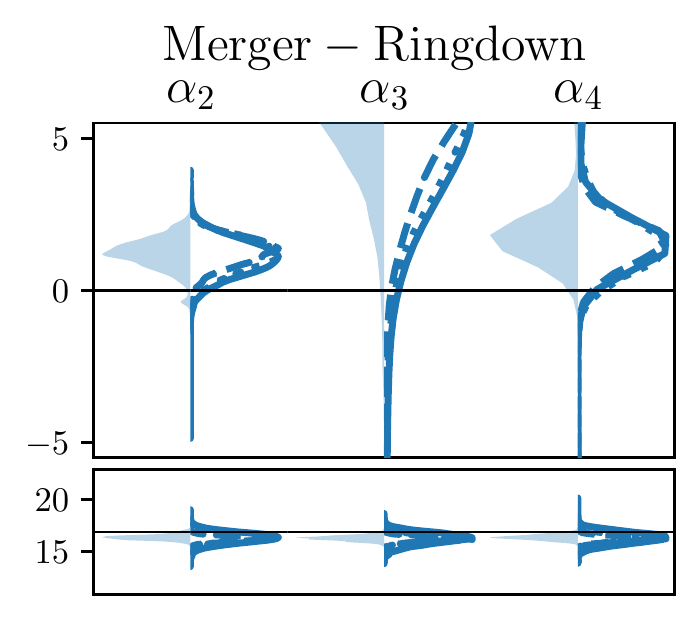}}\\
	\includegraphics[width=\textwidth]{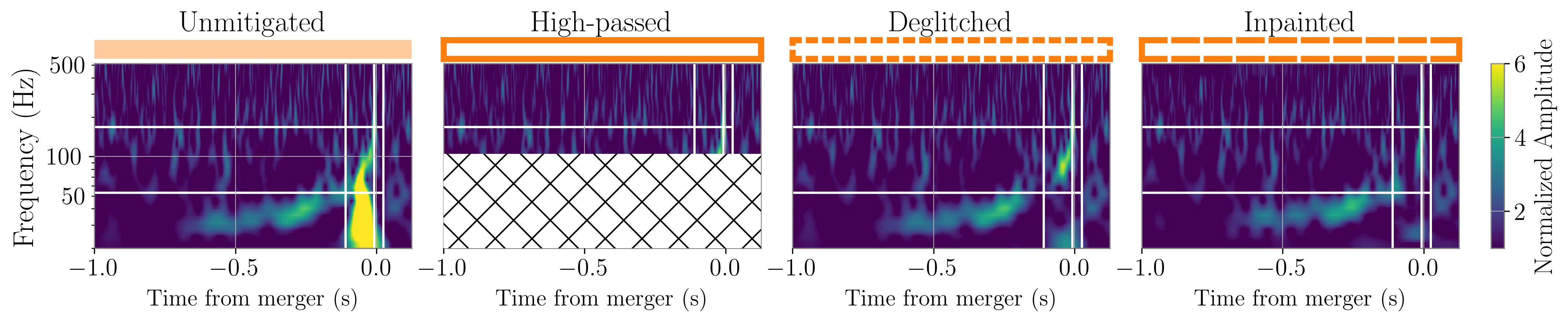}\\[-1.0ex]
	\subfloat[Simulated GW190828\_065509-like signal overlapped with a L1 tomte glitch at intermediate stage in the time domain.]{\includegraphics[height=0.15\textheight]{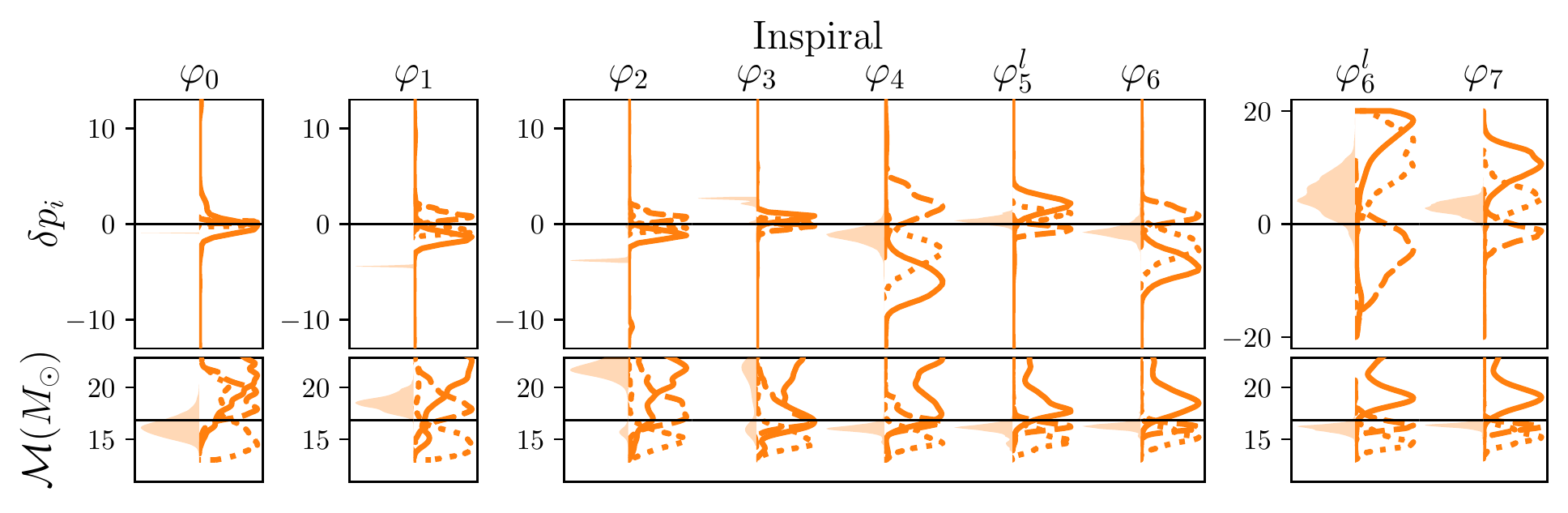}\label{fig:tomte_intermediate}

	\includegraphics[height=0.15\textheight]{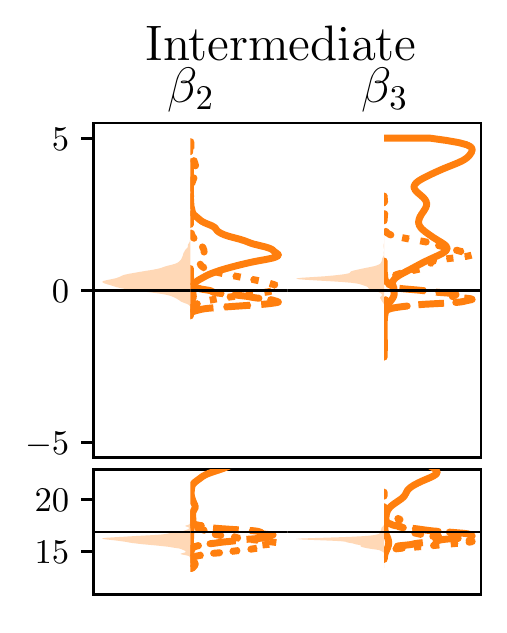}
\includegraphics[height=0.15\textheight]{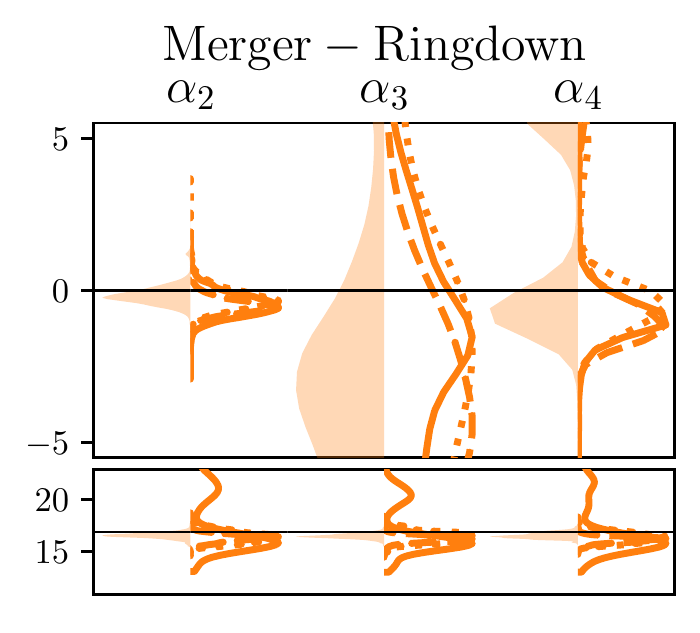}}\\
	\includegraphics[width=\textwidth]{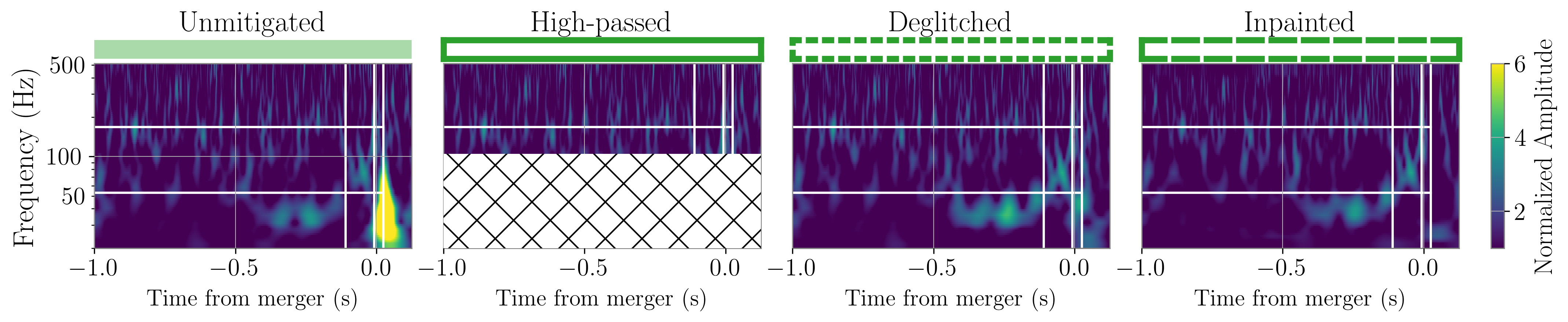}\\[-1.0ex]
	\subfloat[Simulated GW190828\_065509-like signal overlapped with a L1 tomte glitch at merger-ringdown stage in the time domain.]{\includegraphics[height=0.15\textheight]{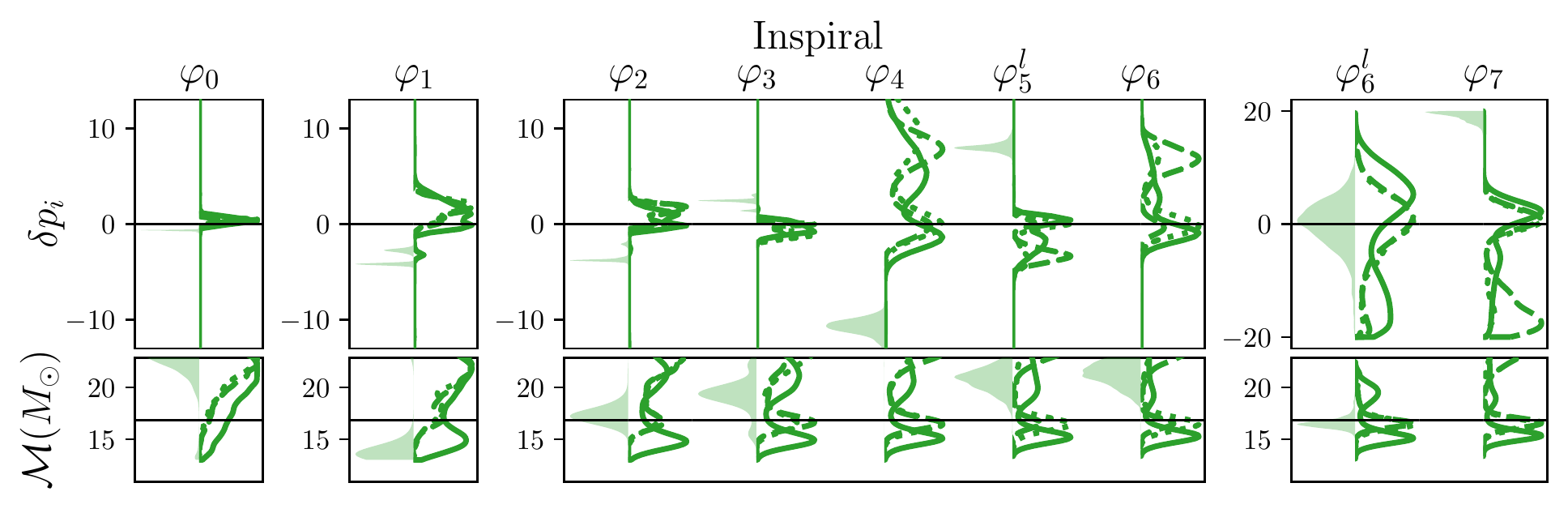}\label{fig:tomte_mr}
	\includegraphics[height=0.15\textheight]{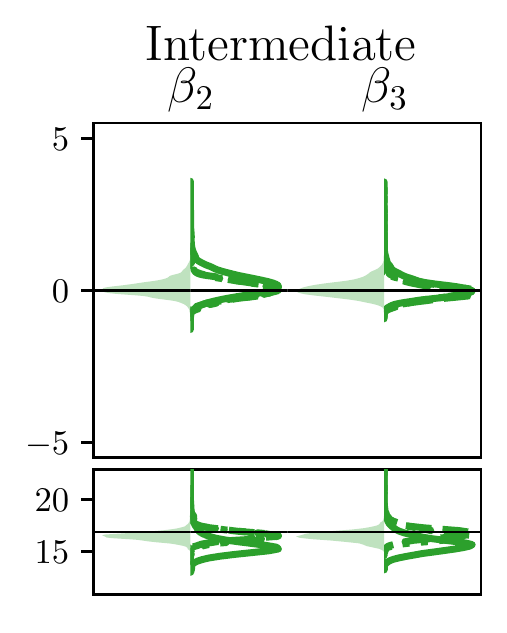}
\includegraphics[height=0.15\textheight]{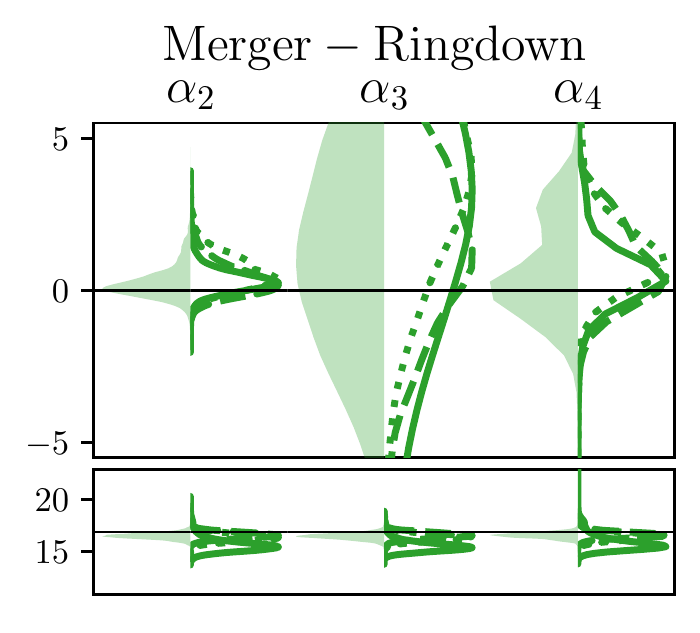}}\\
\caption{Similar to Fig.\ \ref{fig:blip_violins}, but for tomte-glitch-overlapped signals.}
	\label{fig:tomte_violins}
\end{figure*}
Tomte glitches are short-duration, broadband glitches characterized by their triangular shape as seen in time-frequency representations. A $Q$ scan of a tomte glitch is shown on the top right in Fig.\ \ref{fig:glitch}. The sources and coupling of tomte glitches are not well understood.

The simulated GW190828\_065509-like signal is coherently injected into H1, L1 and V1 in a way that a tomte glitch at L1 overlaps with the signal at the inspiral, intermediate, and merger-ringdown stages in the time domain.
Mitigation methods are applied to L1 data, and parametrized tests of GR are performed on the unmitigated and mitigated data. The posteriors of the testing parameters are plotted on the left and right side of each violin plot in Fig.\ \ref{fig:tomte_violins}, respectively, while the unmitigated and mitigated data from L1 are represented by $Q$ scans.

The stages of coalescence where violations of GR are observed show no correlation with those overlapped by the glitch in the time or frequency domain: Despite shifting the time of overlap of the signal with the tomte glitch, which contributes considerable excess power also to the intermediate frequency bands, false violations of GR are observed only in inspiral testing parameters for the unmitigated data samples (left of violins). For example, exclusions of the GR value of 0 at 90\% credibility are observed in lower PN orders such as $\delta\varphi_2$, $\delta\varphi_3$, and $\delta\varphi_4$.

Again, we can observe bias in the inference of chirp mass when the tomte glitch overlaps the simulated signal, such that the posterior distributions peak far away and exclude the injected value at 90\% credibility [e.g., $\delta\varphi_3$ in Fig.\ \ref{fig:tomte_inspiral}, $\delta\varphi_2$ in Fig.\ \ref{fig:tomte_intermediate}, and  $\delta\varphi_4$ in Fig.\ \ref{fig:tomte_mr}]. The independent glitch mitigation methods bring the peak values of the chirp mass close to the injected value, yet multimodal features can still be observed for both chirp mass and inspiral PN parameters due to the degeneracies between them, as demonstrated in Fig.\ \ref{fig:tomte_mr_corner}.

Comparing the unmitigated and mitigated results, both inpainting and glitch model subtraction can reduce the false violations in the lower PN order testing parameters, resulting in strong support for the GR value of 0 in most testing parameters.
Improvements in parametrized tests of GR after removal of the glitch by inpainting and glitch model subtraction suggest that the false violations in the inspiral parameters are attributed to the presence of the tomte glitch, which contributes significant excess power in inspiral frequency bands.

\begin{figure}
	\includegraphics[width=\linewidth]{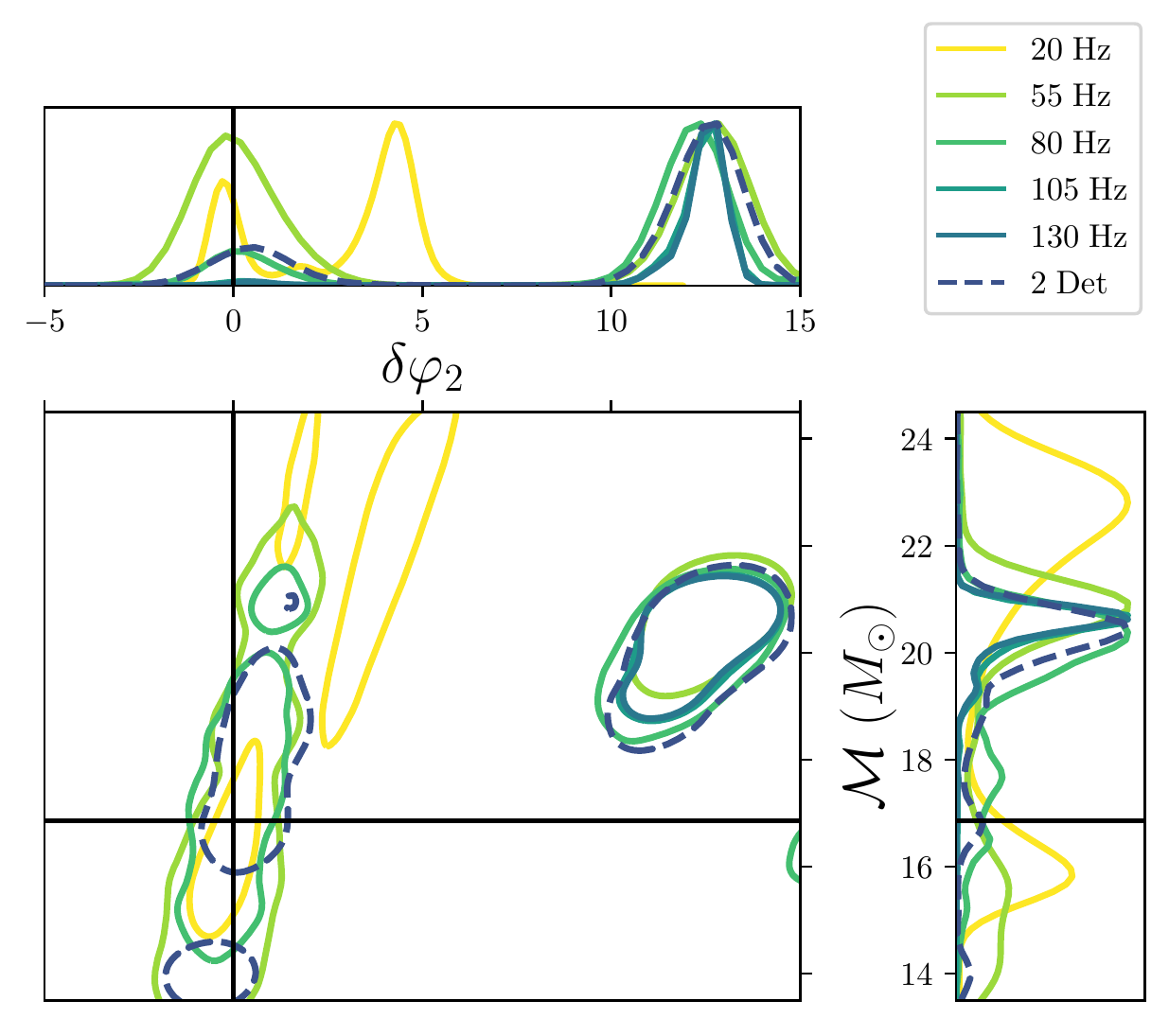}
	\caption{Marginalized distributions of $\delta\varphi_2$ and chirp mass. The contours of the two-dimensional distributions show 90\% credible regions. Solid curves with different colors represent different distributions obtained by high passing the tomte-glitch-overlapped data in L1 shown in Fig.\ \ref{fig:blip_insp} to different frequencies. The dotted curves represent distributions obtained by discarding data from L1 and performing two-detector observation with H1 and V1 data only. The vertical and horizontal black lines denote the GR value of the testing parameter and the injected value of chirp mass, respectively.} 
	\label{fig:highpass}
\end{figure}

Meanwhile, high passing up to 105 Hz is not a robust glitch mitigation method, as false deviations of GR can be amplified [e.g., $\delta\varphi_2$ and $\delta\varphi_3$ in Fig.\ \ref{fig:tomte_inspiral}] or introduced by the mitigation [e.g.\ $\delta\beta_2$ and $\delta\beta_3$ in Fig.\ \ref{fig:tomte_intermediate}]. 
Figure \ref{fig:highpass} shows an example of increasing the high-pass cutoff from the default specifications of 20 Hz for unmitigated samples by increments of 25 or 35 Hz up to 130 Hz and compares the distributions with that obtained by only using data from H1 and V1. It is found that, as more and more signal (glitch) power is discarded by highpassing, the distributions converge to a peak away from the injected value (and away from the biased distributions due to the glitch). It is then found that using only H1 and V1 data in the absence of glitches would recover this bias.
This indicates that further reduction of signal power from the weak signal can lead to bias. While the source of this bias, or its relationship with the bias observed in Fig.\ \ref{fig:gaussian_corner_dalpha2} for $\delta\alpha_2$ in simulated Gaussian noise, are unidentified, this anomalous effect can be easily decoupled from the effects due to glitches by performing independent glitch mitigation methods as illustrated in Figs.\ \ref{fig:tomte_violins} and \ref{fig:highpass}.

Representative figures of the SNR for unmitigated and mitigated samples are obtained in the same way as in Sec.\ \ref{sec:results_blip} and are $13.93^{+0.22}_{-0.36}$ before mitigation, $12.91^{+0.25}_{-0.43}$ after high passing, $13.51^{+0.27}_{-0.36}$ after glitch subtraction, and $13.99^{+0.23}_{-0.38}$ after inpainting.

\begin{figure*}
	\includegraphics[width=\textwidth]{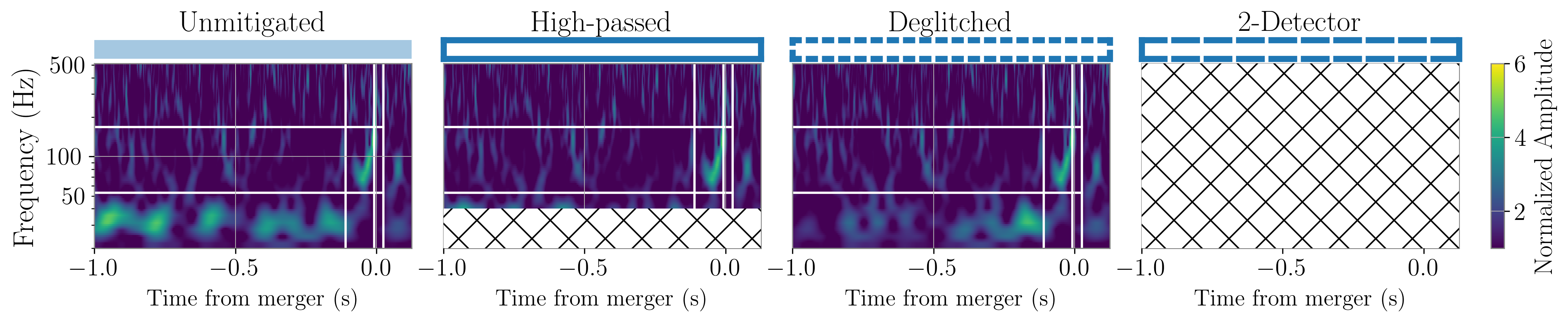}\\[-1.0ex]
	\subfloat[Simulated GW190828\_065509-like signal overlapped with a H1 scattered-light glitch at inspiral stage in the time domain.]{\includegraphics[height=0.15\textheight]{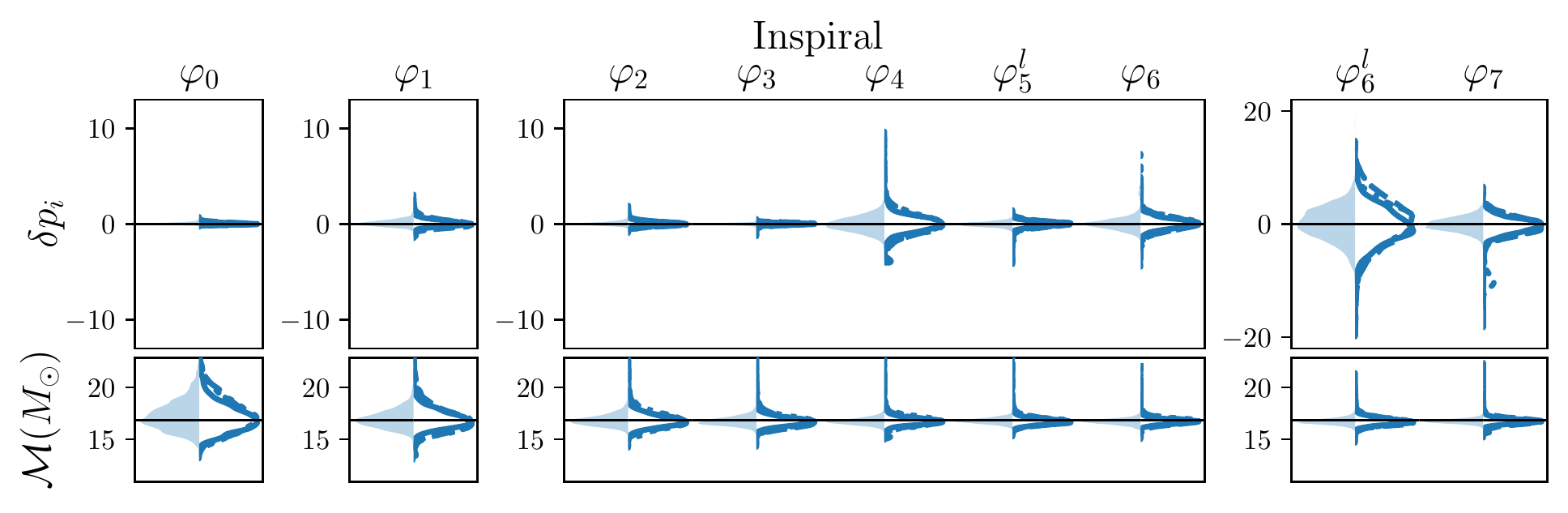}\label{scatter:inspiral}
	\includegraphics[height=0.15\textheight]{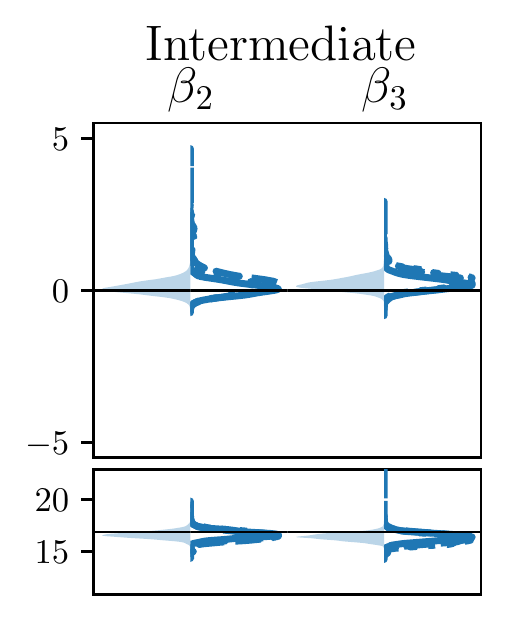}
\includegraphics[height=0.15\textheight]{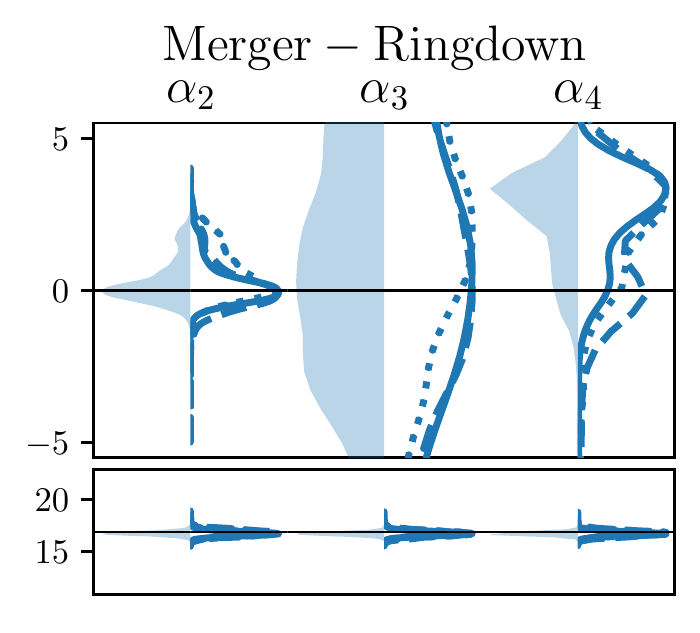}}\\
	\includegraphics[width=\textwidth]{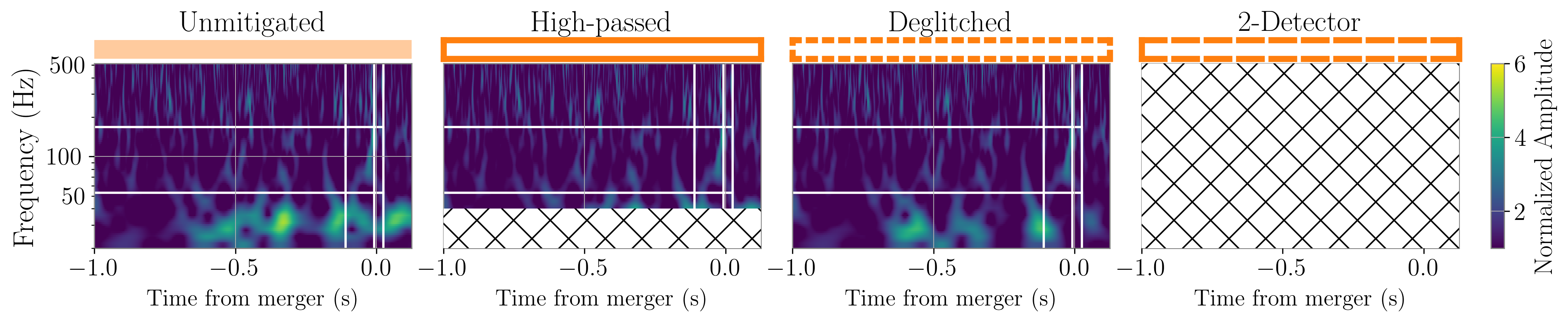}\\[-1.0ex]
	\subfloat[Simulated GW190828\_065509-like signal overlapped with a H1 scattered-light glitch at all stages in the time domain.]{\includegraphics[height=0.15\textheight]{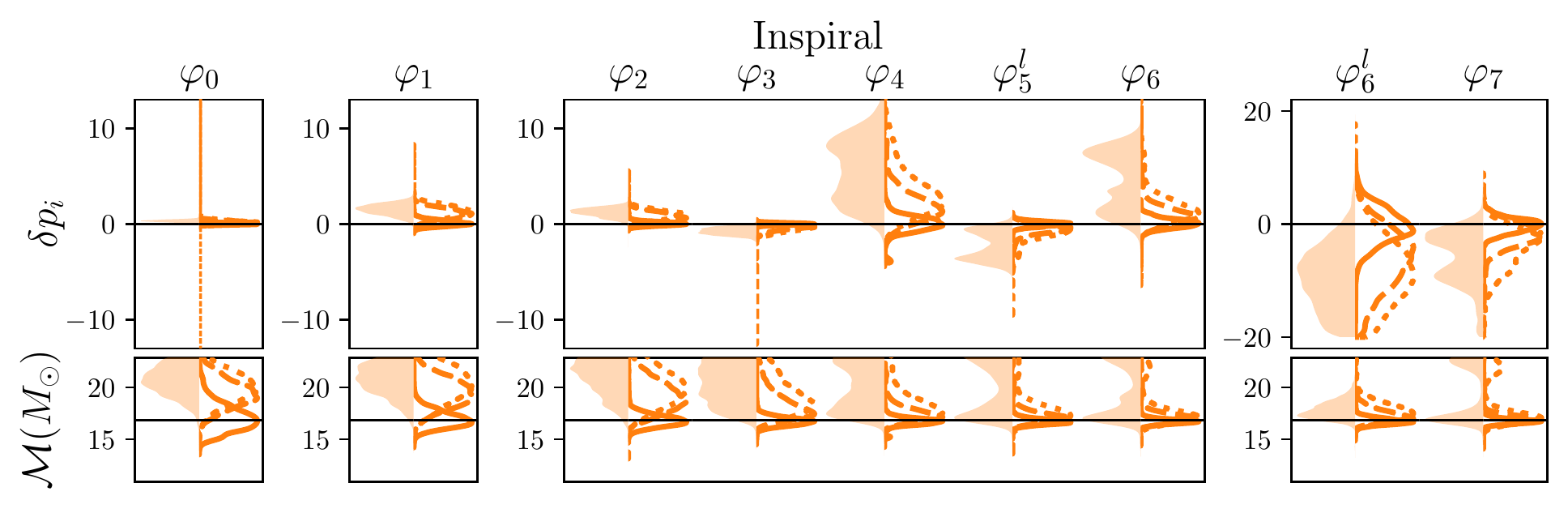}\label{scatter:intermediate}
	\includegraphics[height=0.15\textheight]{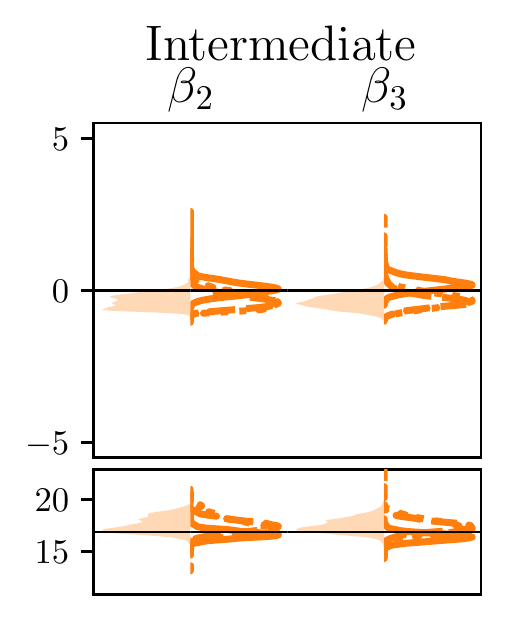}
\includegraphics[height=0.15\textheight]{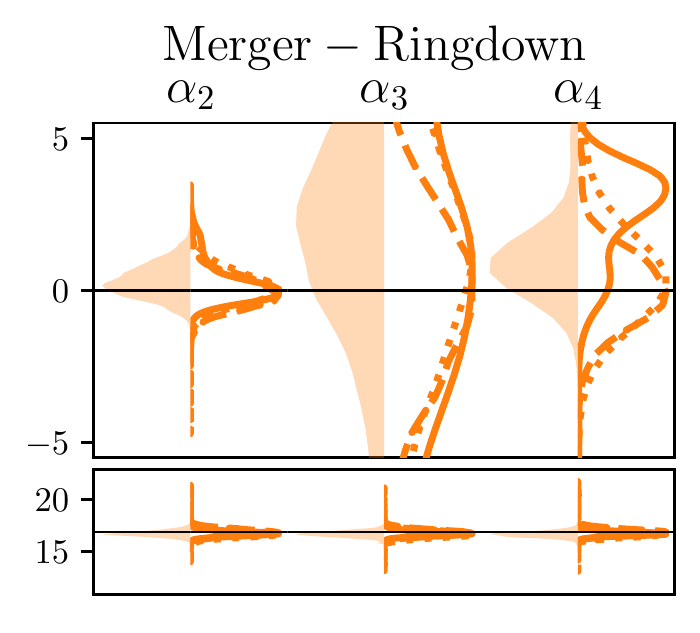}}\\
	\includegraphics[width=\textwidth]{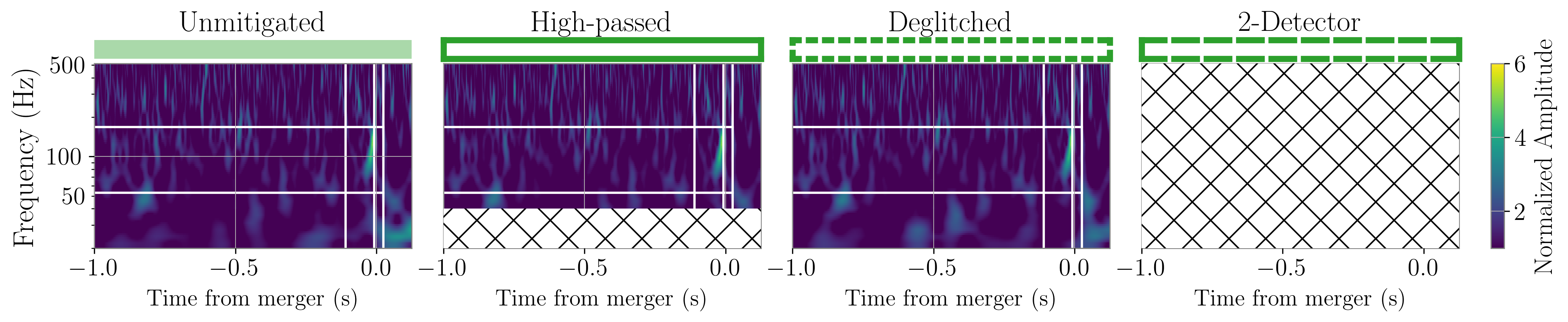}\\[-1.0ex]
	\subfloat[Simulated GW190828\_065509-like signal overlapped with a H1 scattered-light glitch at merger-ringdown stage in the time domain.]{\includegraphics[height=0.15\textheight]{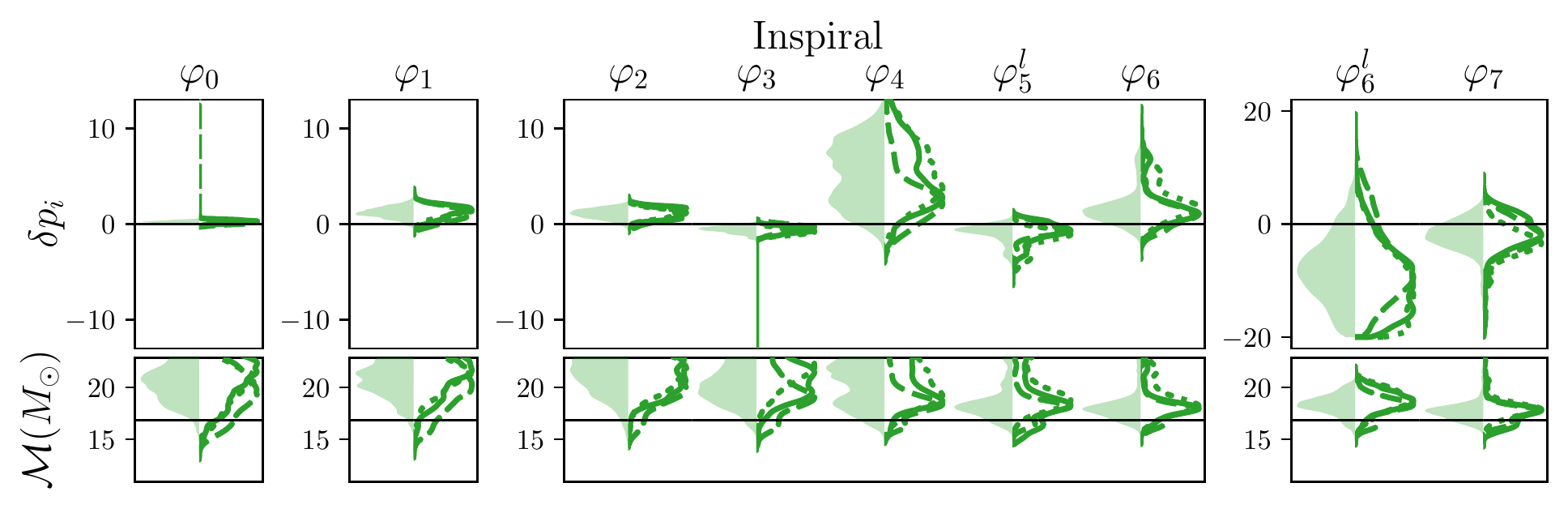}\label{scatter:mr}
	\includegraphics[height=0.15\textheight]{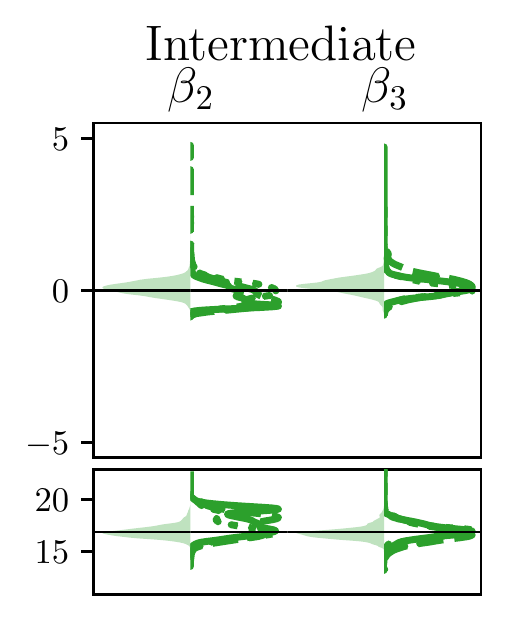}
\includegraphics[height=0.15\textheight]{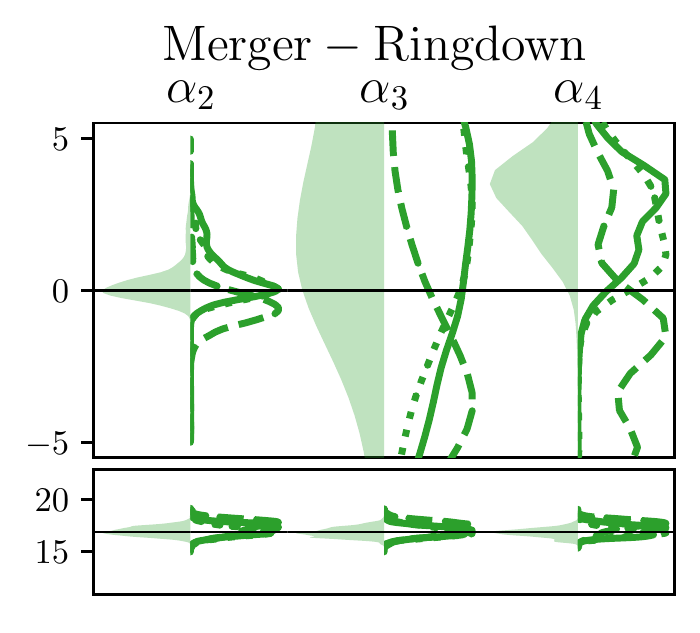}}\\
\caption{Similar to Fig.\ \ref{fig:blip_violins}, but for scattered-light-glitch-overlapped signals and with inpainting replaced with using only data from detectors without glitches.}
	\label{fig:scatter_violins}
\end{figure*}
\subsection{Scattered-light glitch} \label{sec:results_scattered-light}
Scattered-light glitches are produced by laser light scattering out and reentering the main laser beam, and their correlation with seismic motion is well understood \cite{gwtc2}. Scattered-light glitches are characterized by their arch shape as seen in a time-frequency representation such as the bottom $Q$ scan in Fig.\ \ref{fig:glitch} \cite{glitch2}.

%
The simulated GW190828\_065509-like signal is coherently injected into H1, L1 and V1 in a way that a scattered-light glitch at H1 overlaps with the signal at the inspiral, inspiral-intermediate-merger-ringdown, and merger-ringdown stages in the time domain.
Mitigation methods are applied to H1 data, and parametrized tests of GR are performed on the unmitigated and mitigated data. The posteriors of the testing parameters are plotted on the left and right side of each violin plot in Fig.\ \ref{fig:scatter_violins}, respectively, while the unmitigated and mitigated data from H1 are represented by $Q$ scans.

Since the typical timescale of scattered-light glitches ($>$1 s) is large compared to that of BBH-coalescence GW signals (in the LIGO band), it is likely for an overlapping scattered-light glitch to overlap with the entire GW signal.
As such, we choose not to apply time-domain filtering methods, such as inpainting, to the study of scattered-light-glitch-overlapped signals.
To compare the results of the other mitigation methods, we replace inpainting with using data only from the remaining two detectors in which glitches are not present.
However, such a method is not preferred, in general, as much useful information is lost, and is omitted from later discussions of suitable mitigation methods; the example illustrated in Fig.\ \ref{fig:highpass} also suggests that such a method may as well lead to bias.

For all three cases of glitch overlapping, the value of 0 is not excluded at 90\% credibility from the posterior distributions of testing parameters for the unmitigated samples (left of violin plots).
Furthermore, the glitch has no observable effect on parametrized tests of GR when it temporally overlaps with the inspiral stage of the signal [Fig.\ \ref{scatter:inspiral}], as the posterior distributions of all testing parameters for the unmitigated case match with that with the glitch removed in three independent methods. 

In most cases, the posterior distributions of testing parameters obtained by the three independent glitch mitigation methods match closely with each other. This is a good indication that the mitigation methods do not introduce additional effects to the results.
In particular, in the case where the scattered-light glitch overlaps all stages of coalescence in the time domain [Fig.\ \ref{scatter:intermediate}], the glitch removal through high passing to 40 Hz improves the inference of inspiral testing parameters most significantly (see $\delta\varphi_4$, $\delta\varphi_{5l}$, $\delta\varphi_6$,  $\delta\varphi_{6l}$, and  $\delta\varphi_7$.)
However, from the investigation in Fig.\ \ref{fig:highpass}, we would warn against discarding data from the affected detector entirely, as substantial reduction of signal power may lead to bias.

Representative figures of the SNR for unmitigated and mitigated samples are obtained in the same way as in Sec.\ \ref{sec:results_blip}, and are $15.06^{+0.23}_{-0.37}$ before mitigation, $15.04^{+0.24}_{-0.37}$ after high passing, $14.21^{+0.25}_{-0.39}$ after glitch subtraction, and $12.19^{+0.27}_{-0.43}$ after discarding data from H1 where the scattered-light glitch is present.
\section{Conclusion and Outlook}
We overlapped a simulated high-mass-ratio coalescing BBH signal with three glitches from most frequently occurring glitch classes in O3a. We then investigated the effects on parametrized tests of GR of the glitches and their mitigation through bandpass filtering, inpainting, and \texttt{BayesWave} glitch model subtraction.
Although the number of glitches considered in this investigation is not sufficient for us to give quantitative statements about the effects of certain glitch classes or mitigation methods on tests of GR, our analysis covered all stages of BBH coalescence in the time and frequency domain, and we are able to identify the effects case by case by comparing the unmitigated results with that mitigated by independent methods and expected GR results when the noise model is not violated.

No false violations of GR are identified for data samples overlapped with the scattered-light glitch, while false violations are observed for that overlapped with the tomte and blip glitches. For the latter cases, we found no clear correlation between the stages of coalescence in which false violations occurred and those overlapped by the glitch in the time or frequency domain.

Out of the three mitigation methods, we find that inpainting and \texttt{BayesWave} glitch model subtraction consistently reduce false violations of GR, and the results match closely with each other. This indicates that the two methods did not introduce additional effects to parametrized tests  and suggests successful glitch removals. Bandpass filtering, on the other hand, can also reduce false violations in most cases. However, false violations are amplified or new violations are introduced in more than one case after substantial removal of signal power through high passing or discarding data from the glitch-affected detector. We suggest the application of inpainting or \texttt{BayesWave} glitch model subtraction for glitch mitigation, as they are found to be effective even when an extra degree of freedom is involved with the introduction of parametrized deviation to the signal model.

A major improvement on the LIGO detectors is expected to be completed in a few years. The increased sensitivity, in turn, suggests more frequent occurrence of glitches overlapping signals. As mitigating signals overlapped with glitches may become a regularity in the future, a systematic study on the effects of glitches and their mitigation to tests of GR will be crucial to the next generation of GW astronomy. Such a study would likely involve similar methodologies to that presented by this work, applied repeatedly to study different conditions, making this work an important first study on this subject.

\section{Acknowledgments}
J.\ Y.\ L.\ K thanks Derek Davis and Jonah Kanner for insightful discussions. We thank the referee for careful reading of the manuscript; the referee's suggestions have greatly improved our investigation. We thank the National Science Foundation (NSF) and NSF Research Experiences for Undergraduates (REU) Program for supporting the LIGO Summer Undergraduate Research Fellowships (SURF) program.  
The LIGO SURF Program is supported by NSF Grant No.\ PHY-1852081.
Computing resources for this study was provided by the LIGO Laboratory and supported by NSF Grants No.\ PHY-0757058 and No.\ PHY-0823459.
The work described in this paper is partially supported by grants from the Research Grants Council of the Hong Kong (Project No. CUHK 24304317), The Croucher Foundation of Hong Kong and Research Committee of the Chinese University of Hong Kong.
This research has made use of data, software, and/or Web tools obtained from the Gravitational Wave Open Science Center (\href{https://www.gw-openscience.org/}{https://www.gw-openscience.org/} ), a service of LIGO Laboratory, the LIGO Scientific Collaboration and the Virgo Collaboration. LIGO Laboratory and Advanced LIGO are funded by the United States National Science Foundation (NSF) as well as the Science and Technology Facilities Council (STFC) of the United Kingdom, the Max-Planck-Society (MPS), and the State of Niedersachsen, Germany, for support of the construction of Advanced LIGO and construction and operation of the GEO600 detector. Additional support for Advanced LIGO was provided by the Australian Research Council. Virgo is funded, through the European Gravitational Observatory (EGO), by the French Centre National de Recherche Scientifique (CNRS), the Italian Istituto Nazionale di Fisica Nucleare (INFN) and the Dutch Nikhef, with contributions by institutions from Belgium, Germany, Greece, Hungary, Ireland, Japan, Monaco, Poland, Portugal, and Spain.
This paper carries LIGO Document No.\ LIGO-P2100294.

\bibliography{references}
\end{document}